\documentclass[a4paper,11pt]{article}
\pdfoutput=1 

\usepackage{jheppub} 

\usepackage[T1]{fontenc} 
\usepackage{amsmath,amssymb}
\usepackage{graphicx}

\makeatletter

\renewcommand\section{\@startsection{section}{1}{\z@}
                                   {-3.5ex \@plus -1ex \@minus -.2ex}
                                   {2.3ex \@plus .2ex}
                                   {\normalfont\large\bfseries}}
\renewcommand\subsection{\@startsection{subsection}{2}{\z@}
                                   {-3.25ex\@plus -1ex \@minus -.2ex}
                                   {1.5ex \@plus .2ex}
                                   {\normalfont\normalsize\bfseries}}
\renewcommand\subsubsection{\@startsection{subsubsection}{3}{\z@}
                                   {-3.25ex\@plus -1ex \@minus -.2ex}
                                   {1.5ex \@plus .2ex}
                                   {\normalfont\normalsize\bfseries}}
\renewcommand\paragraph{\@startsection{paragraph}{4}{\z@}
                                   {3.25ex \@plus1ex \@minus.2ex}
                                   {-1em}
                                   {\normalfont\normalsize\bfseries}}

\makeatother

\newdimen\tableauside\tableauside=1.0ex
\newdimen\tableaurule\tableaurule=0.4pt
\newdimen\tableaustep
\def\phantomhrule#1{\hbox{\vbox to0pt{\hrule height\tableaurule
width#1\vss}}}
\def\phantomvrule#1{\vbox{\hbox to0pt{\vrule width\tableaurule
height#1\hss}}}
\def\sqr{\vbox{%
  \phantomhrule\tableaustep

\hbox{\phantomvrule\tableaustep\kern\tableaustep\phantomvrule\tableaustep}%
  \hbox{\vbox{\phantomhrule\tableauside}\kern-\tableaurule}}}
\def\squares#1{\hbox{\count0=#1\noindent\loop\sqr
  \advance\count0 by-1 \ifnum\count0>0\repeat}}
\def\tableau#1{\vcenter{\offinterlineskip
  \tableaustep=\tableauside\advance\tableaustep by-\tableaurule
  \kern\normallineskip\hbox
    {\kern\normallineskip\vbox
      {\gettableau#1 0 }%
     \kern\normallineskip\kern\tableaurule}%
  \kern\normallineskip\kern\tableaurule}}
\def\gettableau#1 {\ifnum#1=0\let\next=\null\else
  \squares{#1}\let\next=\gettableau\fi\next}

\makeatletter

\renewcommand\section{\@startsection{section}{1}{\z@}
                                   {-3.5ex \@plus -1ex \@minus -.2ex}
                                   {2.3ex \@plus .2ex}
                                   {\normalfont\large\bfseries}}
\renewcommand\subsection{\@startsection{subsection}{2}{\z@}
                                   {-3.25ex\@plus -1ex \@minus -.2ex}
                                   {1.5ex \@plus .2ex}
                                   {\normalfont\normalsize\bfseries}}
\renewcommand\subsubsection{\@startsection{subsubsection}{3}{\z@}
                                   {-3.25ex\@plus -1ex \@minus -.2ex}
                                   {1.5ex \@plus .2ex}
                                   {\normalfont\normalsize\bfseries}}
\renewcommand\paragraph{\@startsection{paragraph}{4}{\z@}
                                   {3.25ex \@plus1ex \@minus.2ex}
                                   {-1em}
                                   {\normalfont\normalsize\bfseries}}

\makeatother

\newcommand{\id}{\hbox{1\kern-.27em l}}
\newcommand{\ZZ}{\mathbb{Z}}
\newcommand{\RR}{\mathbb{R}}
\newcommand{\half}{ {\textstyle \frac{1}{2}  } }

\newcommand{\al}{\alpha}

\newcommand{\bet}{\beta}

\newcommand{\ka}{\kappa}
\newcommand{\de}{\delta}

\newcommand{\la}{\lambda}

\newcommand{\tha}{\theta}

\newcommand{\Ups}{\Upsilon}

\newcommand{\cN}{\mathcal{N}}

\newcommand{\D}{{\rm d}}

\newcommand{\rar}{\rightarrow}
\newcommand{\emp}{\emptyset}
\newcommand{\non}{\nonumber}

\newcommand{\SO}{\mathrm{SO}}

\newcommand{\Sp}{\mathrm{Sp}}
\newcommand{\su}{\mathrm{su}}
\newcommand{\so}{\mathrm{so}}
\newcommand{\spl}{\mathrm{sp}}

\newcommand{\Spin}{\mathrm{Spin}}

\newtheorem{Theorem}{Rule}

\title{\boldmath  Symbol, Surface operators and  $S$-duality }

\author[b]{ShengLiang Cui}
\author[a,1]{Bao Shou}

\affiliation[b]{College of Physics and Electronic Engineering\\
Hainan Normal University \\
Haikou 571158, China}

\affiliation[a]{Center  of Mathematical  Sciences\\
Zhejiang University \\
Hangzhou 310027,China}

\emailAdd{shlcui@pku.edu.cn}
\emailAdd{bsoul@zju.edu.cn}

\abstract{
We study rigid surface operators in the $\cN=4$ supersymmetric Yang-Mills theories with gauge groups $\SO(n)$ and  $\Sp(2n)$.  Using maps $X_S$ and $Y_S$ between these two theories,  Wyllard made explicit proposals for how the $S$-duality map should act on certain subclasses of surface operators.  We study the maps $X_S$ and $Y_S$ further and simplify   the construction of  symbol invariant of rigid surface operators by a convenient trick. By consistency checks, we recover and extend the $S$-duality maps proposed by Wyllard. We  find  new subclasses of  rigid  surface operators related by $S$-duality. We try to explain the exceptions of  $S$-duality maps.   We also discuss the extension of the techniques used in the $B_n/C_n$ theories to the $D_n$ theories.
}

\begin{document}
\maketitle
\flushbottom

\section{Introduction}

Surface operators  are supported on a two-dimensional submanifold of spacetime, which are  natural generalisations of the Wilson and 't~Hooft operators.  In \cite{Gukov:2006},  Gukov and Witten initiated a study of surface operators in $\cN=4$ super Yang-Mills theories.

$S$-duality for certain subclass of surface operators is discussed  in \cite{Witten:2007}\cite{Wyllard:2009}. The $S$-duality conjecture \cite{Montonen:1977} for the  $\cN = 4$ super  Yang-Mills theories in four dimensions  asserts that the theory with gauge group $G$ and  complexified coupling constant $\tau $
 is equivalent to the theories arising from the transformations $S$ and $T$:
\begin{eqnarray}
S : \; (G, \tau) & \rightarrow & ( G^{L}, - 1 / n_{\mathfrak{g}} \tau)\, , \non \\
T : \; (G, \tau) & \rightarrow & (G, \tau + 1) \,,\non
\end{eqnarray}
where $G^{L}$ denotes the Langlands dual group of $G$.   $n_{\mathfrak{g}}$ is 2 for $F_4$,  3 for $G_2$, and 1 for other semisimple classical groups \cite{Gukov:2006}. For example, the Langlands dual groups of $\Spin(2n{+}1)$ are $\Sp(2n)/\ZZ_2$. And the langlands dual groups of $\SO(2n)$ are  themselves.

In \cite{Gukov:2008},  Gukov and Witten extended their earlier analysis \cite{Witten:2007} of surface operators  in the $\cN=4$ super-Yang-Mills theories. They identified a subclass of surface operators called {\it 'rigid'} surface operators,   preserving  half the supersymmetries.  Rigid surface operators appear not to suffer from  quantum ambiguities, expected to be closed under $S$-duality. There are two types rigid surface operators:  unipotent and semisimple. The rigid semisimple surface operators  are labelled by pairs of  partitions. Unipotent rigid surface operators arise  when one of the  partitions is empty.

In  \cite{Wyllard:2009}, some proposals for the $S$-duality maps related to rigid surface operators in the $B_n$($\SO(2n{+}1)$) and $C_n$($\Sp(2n)$) theories were made. These  proposals involved  all unipotent rigid surface operators as well as certain subclasses of rigid semisimple operators. A problematic mismatch in the total number of rigid surface operators between  the $B_n$ and the $C_n$ theories was pointed out by Wyllard.

In this paper,  we attempt to extend the analysis in \cite{Wyllard:2009}. With no noncentral rigid conjugacy classes in the $A_n$ theory, we do not discuss surface operators in this case. We omit the discussion of the exceptional groups, which are more complicated.     We will focus on theories with gauge groups $\SO(2n)$  and the gauge groups $\Sp(2n) $ whose Langlands dual group is $\SO(2n+1)$. In section \ref{surf},  we review the construction of rigid surface operators given in \cite{Gukov:2008}.  We discuss some mathematical results and definitions as preparation. In section \ref{inv},   we focus on the  invariants of  surface operators which are unchanged under the $S$-duality map.  We review the  \textit{symbol} invariant proposed in \cite{Wyllard:2009}. In particular, we find  a simple rule to determine the contribution to symbol of an even row in a partition.  In section \ref{bcd}, we introduce two maps $X_S$ and $Y_S$ preserving symbol invariant.  Using these maps and the construction of symbols for partitions with only even rows,  we  simplify the computation of symbols for general rigid surface operators in the $B_n$, $C_n$, and $D_n$ theories.

In section \ref{BC},   we review the dual pair of rigid surface operators proposed in \cite{Wyllard:2009}.
By consistency check, we recover  the $S$-duality maps proposed by Wyllard and  find  new subclasses of  rigid  surface operators related by $S$-duality.
As an example,  we tabulate all rigid surface operators and their associated invariants in the $\SO(13)$ and $\Sp(12)$ theories.
We try to explain  exceptions of  $S$-duality maps.
We find that there  are  general characteristics in these proposals and make a conjecture.

In section \ref{Ds},  we  extend the discussion    of the techniques used in the $B_n/C_n$ theories to the $D_n$ theories. We tabulate all rigid surface operators and their associated invariants in the $\SO(12)$ to illustrate  our proposals.

In  appendix A, we summarizes revelent facts.

\section{Surface operators in $\cN=4$ Super-Yang-Mills}\label{surf}
In this section, we  introduce the revelent backgrounds for our discussion. We closely  follow paper \cite{Wyllard:2009} to which we refer the reader for more details.

We consider $\cN=4$ super-Yang-Mills theory on $\mathbb{R}^4$ with coordinates $x^0,x^1,x^2,x^3$. The field content: a gauge field as 1-form, $A_\mu$ ($\mu=0,1,2,3$), four Majorana spinors $\psi^a$ ($a=1,2,3,4$) and six real scalars, $\phi_I$ ($I=1,\ldots,6$). All fields take values in the adjoint representation of the gauge group $G$. Surface operators  are defined by prescribing a certain singularity structure of  fields near the surface on which the operator is supported. We  consider surface operators supported on a $\RR^2$ submanifold $D$ which lie at $x_2=x_3=0$. Since the singularity must be chosen to be compatible with half of  supersymmetries,
 the combinations $A=A_2\, \D x^2 +A_3\, \D x^3$ and $\phi = \phi_2 \,\D x^2 + \phi_3 \,\D x^3$ must obey Hitchin's equations \cite{Gukov:2008}
\begin{eqnarray} \label{hitch}
&&F_A - \phi \wedge \phi = 0 \,, \\
&&\D_A\phi =0\,, \D_A\star A = 0\,.\non
\end{eqnarray}
A surface operator is defined as   a solution to these equations with a prescribed singularity along the surface $\mathbb{R}^2$.

 For the superconformal surface operator, setting $x_2+ix_3 = re^{i\tha}$,  the most general possible rotation-invariant Ansatz for $A$ and $\phi$ is
\begin{eqnarray}
A &=& a(r) \, \D \tha \,, \non \\
\phi &=& -c(r) \, \D \tha + b(r) \frac{\D r }{r}  \,.
\end{eqnarray}
Substituting this Ansatz into Hitchin's equations (\ref{hitch}) and defining $s = -\ln r$ ,  one finds that equations (\ref{hitch}) reduce to Nahm's equations
\begin{eqnarray} \label{nahm}
\frac{\D a}{\D s} &=& [b,c]\,, \non \\
\frac{\D b}{\D s} &=& [c,a] \,,\\
\frac{\D c}{\D s} &=& [a,b] \, \non
\end{eqnarray}
which imply that the constant elements $a$, $b$ and $c$ must commute. Surface operators of this type were treated in \cite{Gukov:2006}.

There is another way to obtain conformally invariant surface operator. Nahm's equations (\ref{nahm}) are solved by
\begin{equation} \label{nahmsol}
a = \frac{T_x}{s + 1/f}\,,\qquad b = \frac{T_z}{s + 1/f}\,,\qquad c = \frac{T_y}{s + 1/f} \,,
\end{equation}
 where $T_x,T_y$ and $T_z$ are elements of the lie algebra $\mathfrak{g}$, spanning a representation  of the  $\su(2)$ Lie algebra. These  $T_i$'s are in  the adjoint representation of the gauge group. The surface operator is actually conformal invariant if the function $f$ allowed to fluctuate.

Besides being defined as the  singular solutions of Hitch's equations, the surface operators can be characterised    as the conjugacy class of the monodromy
\begin{equation}
U = P \exp(\oint \mathcal{A}) \,,
\end{equation}
where $\mathcal{A} = A + i \phi$ and the integration is around a circle near $r=0$.  Following from (\ref{hitch}), it is easy to find that $\mathcal{F} = \D \mathcal{A} + \mathcal{A}\wedge \mathcal{A}=0$, which means that $U$ is independent of   deformations of the integration contour. For the surface operators (\ref{nahmsol}), $U$ becomes
\begin{equation} \label{Uplus}
U= P \exp(\frac{2\pi}{s+1/f} \,\,  T_+ ) \,,
\end{equation}
where $T_+\equiv T_x +i T_y$ is nilpotent,  corresponding to  unipotent surface operator.

There are two types of conjugacy classes in a Lie group: unipotent and semisimple. Semisimple classes   can also lead to  surface operators. For a semisimple element $S$ of the gauge group, we can obtain a surface operator with monodromy $V=SU$, with  the following construction.  Near the surface $D$, we require the fields which are solutions to Nahm's equations satisfy the following restriction
\begin{equation} \label{Scond}
S \Ups(r,\tha) S^{-1} = \Ups(r,\tha+2\pi) \,.
\end{equation}

 From all the surface operators constructed from conjugacy classes, a subclass of surface operators called rigid surface operator is relatively easy  to study. Since they are closed on the $S$-duality.  The rigid surface operators are expected to be superconformal and not to depend on any parameters. A unipotent conjugacy classes is called rigid\footnote{The rigid surface operators  here  correspond to   strongly rigid operators in \cite{Wyllard:2009}. } if its dimension is strictly smaller than that of any nearby orbit. All rigid orbits have been classified \cite{Gukov:2008}\cite{Collingwood:1993}.  A semisimple conjugacy classes $S$  is called rigid if the centraliser  of such class is larger than that of any nearby class. Summary, surface operators are called  rigid if they  based on monodromies of the form $V=SU$, with   $U$ is unipotent and rigid and $S$ is semisimple and rigid.

\subsection{Some mathematical definitions and results}
From the above discussions, we see that a classification of unipotent and semisimple conjugacy classes is needed in order to find the possible surface operators.  In this subsection, we will describe  classification of rigid surface operators in the $B_n$($\SO(2n{+}1)$), $C_n$($\Sp(2n)$) and $D_n$($\SO(2n)$) theories in  detail.

The $T_+$ in  Eq.(\ref{Uplus}) can be described in  block-diagonal basis  as follows
 \begin{equation}
 \label{Ti}
 T_+ = \left( \begin{array}{ccc} T_+^{n_1}  & & \\
                        & \ddots &   \\
                        & & T_+^{n_l}
 \end{array} \right ),
 \end{equation}
where $T_+^{n_k}$ is the `raising' generator of the $n_k$-dimensional irreducible representation of $\su(2)$. For the $B_n$, $C_n$ and $D_n$ theories, there are restrictions on the allowed dimensions of the $\su(2)$ irreps since $T_+$ should belong to the relevant gauge group. From the block-decomposition (\ref{Ti}) we see that unipotent (nilpotent) surface operators  are classified by the restricted partitions.

A partition $\la$ of the positive integer $n$ is a decomposition $\sum_{i=1}^l \la_i = n$  ($\la_1\ge \la_2 \ge \cdots \ge \la_l$).  The integer $l$  is called the length of the partition.
 There is a one-to-one correspondence between partition and Young tableaux. For instance the partition $3^22^31$  corresponds to
\begin{equation}
\tableau{2 5 7}
\end{equation}
For partitions $\la$ and $\ka$,  $\la+\ka$ is the partition with parts $\la_i+\ka_i$.  Young diagrams  occur in a number of branches of mathematics and physics. They  are also useful to construct  the eigenstates of Hamiltonian System \cite{Shou:2011} \cite{{Shou:2014}} \cite{{Shou:2015}}.

Unipotent surface operators in the $B_n$($D_n$) theories are in one-to-one correspondence with partitions of $2n{+}1$($2n$) where all even integers appear an even number of times.    Unipotent surface operators in the $C_n$ theories are in one-to-one correspondence with partitions of $2n$ for which all odd integers appear an even number of times. A partition in the $B_n$ or $D_n$($C_n$) theories is called {\it rigid} if it has no gaps (i.e.~$\la_i-\la_{i+1}\leq1$ for all $i$) and no odd (even) integer appears exactly twice. Rigid partition correspond to rigid surface operator.

For the $B_n$ , $C_n$ and $D_n$ theories,  the  rigid semisimple conjugacy classes $S$  correspond to diagonal matrices with elements $+1$ and $-1$ along the diagonal \cite{Gukov:2008}. The   matrices $S$ break the gauge group to its centraliser at the Lie algebra level as follows
\begin{eqnarray}
\so(2n{+}1) &\rar& \so(2k{+}1)\oplus \so(2n-2k) \,, \non \\
\spl(2n) &\rar& \spl(2k)\oplus \spl(2n-2k) \,, \\
\so(2n) &\rar& \so(2k)\oplus \so(2n-2k) \,. \non
\end{eqnarray}
This implies  that the rigid semisimple surface operators  correspond to pairs of partitions  $(\la';\la'')$  in the $B_n$, $C_n$, and $D_n$ \cite{Gukov:2008}.  In the $B_n$ case, $\la'$ is a rigid  $B_k$ partition and  $\la''$ is a  rigid $D_{n-k}$ partition. For the $C_n$ theories,  $\la'$ is a  rigid $C_k$ partition and  $\la''$ is a  rigid $C_{n-k}$ partition.  For the $D_n$ theories,   $\la'$ is a rigid $D_k$ partition and  $\la''$ is a rigid $D_{n-k}$ partition. In the theories under consideration, the rigid unipotent surface operators can be seen  as a limiting case  $\la''=0$.

There is a close relationship between the pair of partition $(\la';\la'')$  and Weyl group.
 For Weyl groups in the $B_n$ , $C_n$,  and $D_n$ theories  both conjugacy classes and irreducible unitary representations are in one-to one correspondence with ordered pairs of partitions $[\al;\bet]$.  $\al$ is a partition of $n_\al$ and $\bet$ is a partition of $n_{\beta}$, with    $n_\al+n_\bet= n$.
Though the conjugacy classes and unitary representations are parameterised by ordered pair of patitons there is no canonical isomorphism between the two sets.

The Kazhdan-Lusztig  map is a  map from the unipotent conjugacy classes of a simple group to the set of conjugacy classes of the Weyl group.
This map can be extended to the case of rigid semisimple conjugacy classes~\cite{Spaltenstein:1992}.
The Springer correspondence is a injective map from the unipotent conjugacy classes of a simple group to the set of unitary representations of the Weyl group. For the classical groups the above  two maps can be described explicitly in terms of partitions. They  correspond to the invariants  \textit{fingerprint}  and \textit{symbol} of partitions in \cite{Collingwood:1993}, respectively.

We recall  the construction of \textit{symbol} in \cite{Wyllard:2009}.  For a  partition $\lambda$ in the $B_n$  theory, we add $l-k$  to the $k$th part of the partition. The following  table  illustrates this process
$$
\begin{array}{rrrrr}
        \lambda_{k}:           & \lambda_1 & \lambda_{2} &\quad \cdots \quad & \lambda_l\\
  l-k:          & l-1 & l-2 &\quad \cdots \quad & 0\\
 l-k+\lambda_{k}: & l-1+\lambda_1 & l-2+\lambda_2& \quad \cdots \quad & \lambda_l
\end{array}
$$
For the terms in the sequence $ l-k+\lambda_{k}$,  we arrange the odd parts in an increasing sequence $2f_i+1$ from \textbf{right to left} and the even  parts in an increasing sequence $2g_i$ from \textbf{right to left} as follows
$$
\begin{array}{rrrr|rrrr}
2f_i+1:  & \quad \cdots \quad & 2f_2+1 & 2f_1+1 &  \quad  2g_i:     & \quad \cdots \quad & 2g_2  &  2g_1\\
 f_i:    & \quad \cdots \quad & f_2    & f_1    &   \quad  g_i:       & \quad \cdots \quad & g_2   &  g_1
\end{array}
$$
Next we calculate the terms $\al_i = f_i-i+1$ and $\bet_i = g_i-i+1$.
Finally we  write the {\it symbol} as follows \footnote{According to the  computation of symbol, the entries in the symbol corresponding to $\lambda_1,\cdots, \lambda_i$ are independent of that corresponding to $\lambda_{i+1},\cdots, \lambda_l$. }
\begin{equation} \label{symbol}
\left(\begin{array}{@{}c@{}c@{}c@{}c@{}c@{}c@{}c@{}} \al_1 &&\al_2&&\al_3&& \cdots \\ &\bet_1 && \bet_2 && \cdots  & \end{array} \right).
\end{equation}
\begin{flushleft}
  \textbf{Example:} For the $B_{10}$ partition $\la=3^3\,2^4\,1^4$,  the symbol is
\begin{equation} \label{exs}
\left(\begin{array}{@{}c@{}c@{}c@{}c@{}c@{}c@{}c@{}c@{}c@{}c@{}c@{}} 0&&0&&1&&1&&1&&1 \\ & 1 && 1 && 1 &&1&&2& \end{array} \right).
\end{equation}
We can view the two rows of the  symbol as two partitions $[1^4,21^4]$ which is the pair of partitions corresponding to a unitary representation of the Weyl group.
\end{flushleft}

For the $C_n$ case, we need  append an extra $0$ as the last part of the partition if the length of the partition is even.  $f_i$ and $g_i$ are constructed  as in the $B_n$ case. Then   we calculate the  terms $\al_i = g_i-i+1$ and the terms $\bet_i = f_i-i+1$.
\begin{flushleft}
  \textbf{Example:} For the  $C_{10}$ partition $\la=3^2\,2^6\,1^2$,  the symbol is
\begin{equation}
\left(\begin{array}{@{}c@{}c@{}c@{}c@{}c@{}c@{}c@{}c@{}c@{}c@{}c@{}} 0&&0&&1&&1&&1&&1 \\ & 1 && 1 && 1 &&1&&2& \end{array} \right)\non
\end{equation}
which is the same as (\ref{exs})\footnote{In fact, these two partitions are related by the map $X_S$ introduced in section 4.1.}.
\end{flushleft}

For the $D_n$ theory, we calculate $f_i$ and $g_i$ exactly as in the $B_n$ case.  Then   we calculate the  terms $\al_i = g_i-i+1$ and  $\bet_i = f_i-i+1$. The number of $f_i$ and $g_i$ are  equal in this case.
Since the  rigid partitions of the $D_n$ theory which always have at least one part equal to 1,  the last entry of the sequence $\bet_i$ is zero. It is     omitted finally.

\section{Invariants of surface operators} \label{inv}
Invariants of the surface operators  do not change under the $S$-duality map \cite{Gukov:2008}\cite{Wyllard:2009}. The dimension $d$ of associated partition is the most basic invariant of a rigid surface operator. It is calculated as follows \cite{Gukov:2008}\cite{Collingwood:1993}:
\begin{eqnarray}
B_n: & d = 2n^2 + n -\half \sum_{k} (s_k')^2 -  \half \sum_{k} (s_k'')^2
+ \half \sum_{k\;\mathrm{odd}} r_k'+ \half \sum_{k\;\mathrm{odd}} r_k'' \,,\non \\
C_n: & d = 2n^2 + n -\half \sum_{k} (s_k')^2 -  \half \sum_{k} (s_k'')^2
- \half \sum_{k\;\mathrm{odd}} r_k'- \half \sum_{k\;\mathrm{odd}} r_k''\,, \\
D_n: & d =2n^2 - n -\half \sum_{k} (s_k')^2 -  \half \sum_{k} (s_k'')^2 \non
+ \half \sum_{k\;\mathrm{odd}} r_k'+ \half \sum_{k\;\mathrm{odd}} r_k''\,,
\end{eqnarray}
where $s'_k$ denotes the number of parts of $\la'$'s that are larger than or equal to $k$. And $r_k'$ denotes the number of parts of $\la'$ that are equal to $k$. Similarly, we define $s_k''$ and $r_k''$ corresponding to $\la''$.

The invariant {\it symbol}  is based on the Springer correspondence which  can be extended  to rigid semisimple conjugacy classes. One can construct the symbol of this rigid semisimple surface operator by calculating the symbols for both $\la'$ and $\la''$,  then add the entries that are `in the same place' of these two results.  An example illustrates the addition rule:
\begin{equation} \label{symboladd}
\left(\begin{array}{@{}c@{}c@{}c@{}c@{}c@{}c@{}c@{}c@{}c@{}c@{}c@{}c@{}c@{}} 0&&0&&0&&0&&0&&1&&1 \\ & 1 && 1 && 1 &&1&&1&&2 & \end{array} \right) +
 \left(\begin{array}{@{}c@{}c@{}c@{}c@{}c@{}c@{}c@{}c@{}c@{}c@{}c@{}} 0&&0&&0&&1&&1&&1 \\ & 1 && 1 &&1&&1&&1 & \end{array} \right)=
\left(\begin{array}{@{}c@{}c@{}c@{}c@{}c@{}c@{}c@{}c@{}c@{}c@{}c@{}c@{}c@{}} 0&&0&&0&&0&&1&&2&&2 \\ & 1 && 2 && 2 &&2&&2&&3 & \end{array} \right).
\end{equation}

There is another invariant  {\it fingerprint } constructed from $(\la';\la'')$ via the Kazhdan-Lusztig map. This invariant is a pair of partitions $[\al;\bet]$ associated with  the Weyl group conjugacy class.
 It is checked that the symbol of a rigid surface operator contains the same amount of information as the fingerprint \cite{Wyllard:2009}. Compared  with the fingerprint invariant, the symbol is much easier to be calculated and  more    convenient   to  find the $S$-duality maps of surface operators.

In \cite{Witten:2007}, it was pointed that two discrete quantum numbers 'center' and 'topology'  are interchanged under $S$-duality. A surface operator can detect topology then its dual should detect the centre and vice versa. However, there are some puzzles using these  discrete quantum numbers to find duality pair   \cite{Wyllard:2009}.
There is another problem that the generating functions for the total number of rigid surface operators show that the number of rigid surface operators in the $B_n$ theory is larger than that   in the $C_n$ theory \cite{Wyllard:2009}, which  was first observed in the $B_4$/$C_4$ theories \cite{Gukov:2008}.

In this paper,  we  ignore these problems  for the moment. We focus on the symbol invariant to  identify certain subsets of rigid surface operators and make proposals for how the $S$-duality map should acts on surface operators. Hopefully, our constructions will be helpful in making new insight to the surface operator.

\subsection{Construction of symbol of partitions with only even rows}
In this section, we propose  rules to compute symbol of  partitions with only even rows.  Viewing    the contribution to the symbol of each row independently, we find distinct  regular patterns of  contributions to symbol of even rows.

We derive these calculation  rules  through  examples. First, we calculate  the  symbol of an old row $1^{2l+1}$ in the  $B_n$ theory which  can be seen as the first row of  a partition.  According to the definition  of symbol, we have
\begin{equation}\label{b1}
  \begin{array}{rrrrr}
        \lambda_{k}:           & 1 & 1 &\cdots & 1\\
  2l+1-k:          &  2l & 2l-1 &\cdots & 0\\
 2l+1-k+\lambda_{k}: &  2l+1 & 2l&\cdots & 1
\end{array}
\end{equation}
which lead to the following sequences  increasing from \textbf{right to left}
$$\begin{array}{c|c}
\begin{array}{rrrrr}
  2f_i+1:          & 2l+1 & 2l-1 &\cdots & 1\\
 f_i: & l& l-1&\cdots & 0
\end{array}
& \quad
\begin{array}{rrrrr}
  2g_i:    & 2l & 2l-2 &\cdots & 2\\
 g_i: & l & l-1 &\cdots & 1
 \end{array} \end{array}
$$
Thus we get two sequences  $\alpha_i =f_i-i+1: \, 0,\, 0,\, \dots , \, 0 $ and  $  \beta_i =g_i-i+1:\,  1, \, 1 ,\,\dots\, ,\,1$.
So the symbol   of $1^{2l+1}$ is
\begin{equation} \label{s1}
 \Bigg(\!\!\!\begin{array}{c}\overbrace{0 \;\;0\cdots 0}^{l+1} \\
\;\underbrace{1 \cdots 1}_l                        \ \end{array} \Bigg).
\end{equation}

Now we add an even row with $2m$($m<l)$ boxes, which lead to a partition $2^{2m}\,1^{2l+1-2m}$ in the $B_n$ theory.  We calculate the symbol as follows
$$
\begin{array}{rrrrrrrrr}
        \lambda_{k}:           & 2 & 2 &\cdots & 2  & 1 &\cdots & 1\\
  2l+1-k:          & 2l & 2l-1 &\cdots & 2l-2m+1 & 2l-2m &\cdots & 0\\
 2l+1-k+\lambda_{k}: & 2l+2 & 2l+1&\cdots & 2l-2m+3 & 2l-2m+1 &\cdots & 0.
\end{array}
$$
Compared  with (\ref{b1}), the first $2m$ terms in the sequence  $2l+1-k$  is added one  while the others are  unchanged. Then we have
$$
\begin{array}{rrrrrrrrr}
  2f_i+1:          & 2l+1 & 2l-1 &\cdots & 2l-2m+3 & 2l-2m+1 &\cdots & 1\\
 f_i: & l& l-1&\cdots & l-m+1 & l-m &\cdots & 0.
\end{array}
$$
Compared with the sequences $f_i$ corresponding to the first row, nothing is changed.  However,  each term of the  sequences  $g_i$   is added one as follows
$$
\begin{array}{rrrrrrrrr}
  2g_i:    & 2l+2 & 2l &\cdots & 2l-2m+4 & 2l-2m &\cdots & 2\\
 g_i: & l+1 & l &\cdots & l-m+2 & l-m &\cdots & 1.
\end{array}
$$
Then the entries of the top row and  bottom row in the symbol are
$$
\begin{array}{rrrrrrrr}
               \alpha_i =f_i-i+1:   & 0,& 0,& \dots, & 0,& 0,& \dots  ,& 0 \\
                \beta_i =g_i-i+1:  & 2, & 2 ,& \dots ,& 2, & 1 ,& \dots ,& 1.
\end{array}
$$
So the symbol of these two rows $2^{2m}\,1^{2l+1-2m}$ is
\begin{equation}
 \Bigg(\!\!\!\begin{array}{c} \overbrace{0 \;\;0\cdots 0 \;\;0\cdots 0 }^{l+1}\\
\;1 \cdots 1 \;\;\underbrace{2 \cdots 2}_{m}                       \ \end{array} \Bigg).
\end{equation}
Compared to formula (\ref{s1}),  the second  row    contribute  to the symbol as follows
\begin{eqnarray}
\Bigg(\!\!\!\begin{array}{c} \overbrace{0 \;\; 0\cdots 0 \;\; {0\cdots0}}^{l+1}  \\
\, 0\cdots 0\;\;  \underbrace{1\cdots 1}_{m}\end{array} \!\!\!\Bigg)
 \end{eqnarray}
Formally,  the  symbol of $2^{2m}\,1^{2l+1-2m}$  can be seen as  the sum of the contribution of each row
\begin{eqnarray}\label{s0}
\Bigg(\!\!\!\begin{array}{c} 0 \;\; 0\cdots 0 \;\; {0\cdots0}  \\
\, 1\cdots 1\;\;  \underbrace{2\cdots 2}_{m}\end{array} \!\!\!\Bigg)=\Bigg(\!\!\!\begin{array}{c} 0 \;\; 0\cdots 0 \;\; {0\cdots0}  \\
\underbrace{\, 1\cdots 1\;\;  1\cdots 1}_{l}\end{array} \!\!\!\Bigg)
+
\Bigg(\!\!\!\begin{array}{c} 0 \;\; 0\cdots 0 \;\; {0\cdots0}  \\
\, 0\cdots 0\;\;  \underbrace{1\cdots 1}_{m}\end{array} \!\!\!\Bigg)
.
 \end{eqnarray}

Next, if we add anther even   row  of $2n $ boxes($n< m$) to the partition, the new partition  is $3^{2n}\,2^{2m-2n}\,1^{2l+1-2m}$.
We calculate the symbol of this partition as follows
$$
\begin{array}{cccccccccc}
    3  & \cdots & 3 & 2 &\cdots & 2  & 1 &\cdots & 1\\
 2l & \cdots & 2l-2n+1 & 2l-2n &\cdots & 2l-2m+1 & 2l-2m &\cdots & 0\\
 2l+3 & \cdots & 2l-2n+4 & 2l-2n+2 & \cdots & 2l-2m+3 & 2l-2m+1 &\cdots & 1
\end{array}
$$
The above three rows correspond to sequences   $\lambda_{k}, 2l+1-k, 2l+1-k+\lambda_{k}$, respectively.
And the following three rows correspond to sequences  $ 2f_i+1,  f_i, \alpha_i $, respectively.
$$
\begin{array}{ccccccccc}
2l+3 &\cdots & 2l-2n+4+1  & 2l-2n+1 &\cdots & 2l-2m+3 & 2l-2m+1 &\cdots & 1\\
l+1&\cdots & l-n+2 & l-n & \cdots & l-m+1 & l-m &\cdots & 0\\
1 &\dots & 1 &0  &\dots & 0 &0& \dots  & 0
\end{array}
$$
We get the entries $\beta_i$ of the  bottom row  of the symbol as follows
$$\begin{array}{ccccccccccc}
  2g_i:    & 2l+2 &\cdots &2l-2n+4 & 2l-2n+2 &\cdots & 2l-2m+4 & 2l-2m &\cdots & 2\\
 g_i: & l +1 &\cdots &l-n+2 & l-n+1 & \cdots & l-m+2 & l-m &\cdots & 1\\
 \beta_i :  & 2 & \cdots & 2 & 2 & \dots & 2 & 1 & \dots & 1
\end{array}
$$
Thus the symbol of the partition $3^{2n}\,2^{2m-2n}\,1^{2l+1-2m}$ is
\begin{equation}\
 \Bigg(\!\!\!\begin{array}{c}0 \;\;0\cdots\cdots 0 \;\; \overbrace{1\cdots 1}^{n} \\
\;1 \cdots 1 \;\;\underbrace{2  \cdots\cdots 2}_{m}                       \ \end{array} \Bigg) \non
\end{equation}
 which can be  formally   decomposed   into the sum of   the contribution of each row
\begin{eqnarray}
&&\Bigg(\!\!\!\begin{array}{c}0 \;\;0\cdots\cdots 0 \;\; \overbrace{1\cdots 1}^{n} \\
\;1 \cdots 1 \;\;\underbrace{2  \cdots\cdots 2}_{m}  \ \end{array} \Bigg)\non\\
&&=
\Bigg(\!\!\!\begin{array}{c}0 \;\;0\cdots\cdots\cdots \cdots 0 \\
\;1 \cdots 1 \;\;\underbrace{2  \cdots\cdots 2}_{m}   \ \end{array} \Bigg)
+
\Bigg(\!\!\!\begin{array}{c} 0 \;\; 0\cdots 0 \;\; \overbrace{1\cdots1}^n  \\
\, 0\cdots 0\;\; 0\cdots 0 \end{array} \!\!\!\Bigg)
\\&&=
\Bigg(\!\!\!\begin{array}{c}0 \;\;0\cdots\cdots\cdots\cdots 0 \\
\;1 \cdots \cdots\cdots\cdots 1  \ \end{array} \Bigg)
+
\Bigg(\!\!\!\begin{array}{c}0 \;\;0\cdots\cdots\cdots\cdots 0\\
\;0 \cdots 0 \;\;\underbrace{ 1 \cdots\cdots 1}_{m}   \ \end{array} \Bigg)
+
\Bigg(\!\!\!\begin{array}{c} 0 \;\; 0\cdots 0 \;\; \overbrace{1\cdots1}^n  \\
\, 0\cdots 0\;\; 0\cdots 0\end{array} \!\!\!\Bigg) \non
 \end{eqnarray}
Compared with (\ref{s0}), the  contribution  to symbol of the  second  row added to the partition  is
\begin{eqnarray}
\Bigg(\!\!\!\begin{array}{c} 0 \;\; 0\cdots 0 \;\; \overbrace{1\cdots1}^n  \\
\, 0\cdots 0\;\;  {0\cdots 0}\end{array} \!\!\!\Bigg) \non
 \end{eqnarray}
where the number of '1'  is one half of the length of the   row.

According to the  above discussions,  we claim that the contributions to symbol of  even rows have  formal additivity. Since the first row  of a partition in the  $B_n$ theory is odd,   we   consider partitions with only even rows in  the $C_n$ and $D_n$ theories.

\begin{Theorem}{\label{rule}}
\footnote{Addition of  an odd row to a partition leads to  a partition  not in the same theory.  We can  determine the contribution to symbol of each old row  formally,   but there are no simple addition rules for the contribution to symbol\cite{Shou 2}\cite{Shou 3}.}
For a partition with only even rows in  the $C_n$ and $D_n$ theories, the  row with   $2m $ boxes   contribute $m $  '1'  in sequence  from  right to left  in the same   row of symbol,  while other entries of the symbol are  '0'. The contribution to symbol of the adjoining even rows  are formed in the same  way, except that   '1's  occupy another row of symbol.
\end{Theorem}

To  construct  symbol by this rule, the contribution to symbol of the first row is needed as an initial  condition.
In the appendix, we prove  that the longest two rows of a rigid $C_n$ partition  are either  even or odd. This pairwise pattern then continues.  If the first row has $2l$ boxes,  its contribution  to symbol can be calculated as  (\ref{s1})
\begin{equation} \label{C1}
 \Bigg(\!\!\!\begin{array}{c} \overbrace{0\;\;0\cdots 0}^{l+1}  \\
 \;\;\;\underbrace{1 \cdots 1}_{l} \ \end{array} \!\!\!\Bigg).
\end{equation}
For the $D_n$ theory,  we can prove that the longest row of a  partition always contains even number of boxes. And the following two rows are either both of odd length or both of even length. This pairwise pattern then continues. The contribution  to symbol of  the first row with $2l$ boxes is
\begin{equation} \label{D1}
 \Bigg(\!\!\!\begin{array}{c} \overbrace{1\;\;1\cdots 1}^{l}  \\
 \;\;\;\underbrace{0\cdots 0}_{l-1} \ \end{array} \!\!\!\Bigg).
\end{equation}
\begin{flushleft}
\textbf{Example:}  partition  $3^22^21^2$ in the $D_n$ theory,
\end{flushleft}
\begin{equation}
 \tableau{2 4 6}
\end{equation}
According to \textbf{Rule} \ref{rule},  the symbol is
\begin{equation} \label{1}
\sigma_{(3^22^21^2)}^{D}=\left(\begin{array}{@{}c@{}c@{}c@{}c@{}c@{}c@{}}
1&&1&&1 \\ &0&&0& \end{array} \right) +
\left(\begin{array}{@{}c@{}c@{}c@{}c@{}c@{}c@{}} 0&&0&&0 \\ &1&&1& \end{array} \right)+
\left(\begin{array}{@{}c@{}c@{}c@{}c@{}c@{}c@{}} 0&&0&&1 \\ &0&&0& \end{array} \right),
\end{equation}
where the superscript $D$ indicates  it is a  partition  in the $D_n$ theory.

In the  end, we would like to   point out  that the contribution to symbol of the odd rows and  even rows of a partition  are independent of each other.   We illustrate this fact through an example.
We add   three equal rows $3^{2m}$  to a row with $2l+1$ ($l>m$) boxes in the $B_n$ theory, with a decomposition  $3^{2m}=2^{2m}+1^{2m}$.  The decomposition  means  adding  the two  rows $2^{2m}$  firstly then adding the row $1^{2m}$.
From the calculation of   symbol, the rows  $2^{2m}$ will not alter the parity of the sequences $(2l+1)-k+\lambda_{k}$, but do   turn  $f_i$ to $f_i+1$  and $g_i$ to $g_i+1$. Thus its contribution  to symbol  is
\begin{equation}
\Bigg(\!\!\!\begin{array}{c} 0 \;\; 0\cdots 0 \;\; \overbrace{1\cdots1}^m  \\
\, 0\cdots 0\;\;  \underbrace{1\cdots 1}_{m}\end{array} \!\!\!\Bigg). \non
\end{equation}
According to  the  \textbf{Rule} \ref{rule}, the third   row  $1^{2m}$  contribute  to  symbol as follows
\begin{equation}
\Bigg(\!\!\!\begin{array}{c} 0 \;\; 0\cdots 0 \;\; \overbrace{0\cdots 0}^m  \\
\, 0\cdots 0\;\;  \underbrace{1\cdots 1}_{m}\end{array} \!\!\!\Bigg) \non
\end{equation}
which  is equal to the contribution to symbol calculated by the formula (\ref{spesym}) when it is   added to the  row $1^{2l+1}$ firstly.
The  contribution to symbol   will not be changed when one replace rows $2^{2m}$  by another  two rows with the same parity,  since  the entries in the symbol corresponding to $\lambda_1,\cdots, \lambda_i$ are independent of that corresponding to $\lambda_{i+1},\cdots, \lambda_l$.
 This  property and the fact that even or old rows occur pair following the first row in the $B_n$ theory are crucial in searching  the $S$-duality  pairs.

\section{Construction  of  symbol of rigid partitions}\label{bcd}
In the  previous  section, we found an addition     rule  to calculate   symbol  of a partition with only even rows   by summing  the contributions  of each row independently. In this section,  we introduce two   maps $X_S$ and $Y_S$  preserving symbol,  which translate  partition with only odd rows   to  partition with only even rows.   Combining these  maps and employing the addition rule,    we can calculate the symbol of a partition in a simple manner.

\subsection{Map $X_S$}
First, we introduce some concepts. The transpose  partition  $\lambda^t$ is  obtained by interchanging the roles of the rows and columns of the Young tableau corresponding to $\lambda$. For instance
\begin{equation}
\left(\tableau{2 4 5} \right)^t \quad = \quad  \tableau{1 2 2 3 3}
\end{equation}
A partition $\la$ is called {\it special} if its transpose partitions  satisfy  following condition
\begin{eqnarray}
B_n: &\quad \la^t &\;\mbox{is orthogonal} \,,\non \\
C_n: &\quad \la^t &\;\mbox{is symplectic} \,, \\
D_n: &\quad \la^t &\;\mbox{is symplectic} \,.\non
\end{eqnarray}
It is easy to find that all rows of a rigid special partition are  odd in the $B_n$ theory. On the contrary, all rows in the Young tableau corresponding to a rigid special partition are  even in  the $C_n$ and $D_n$ cases.

In \cite{Gukov:2008}, it is pointed  out that the special rigid unipotent surface operators in the $B_n$ and $C_n$ theories are related by $S$-duality. The proposed $S$-duality map $X_S$  performs  in the following way
\begin{eqnarray} \label{XS}
X_S:&& m^{2n_m+1}\, (m-1)^{2n_{m-1}}\, (m-2)^{2n_{m-2}  } \cdots 2^{n_2} \, 1^{2n_1}  \non \\   & \mapsto&
m^{2n_m}\, (m-1)^{2n_{m-1} +2 }\, (m-2)^{2n_{m-2} - 2 } \cdots 2^{n_2+2} \, 1^{2n_1-2}\,.
\end{eqnarray}
Here $m$ has to be odd such that the partition on the left  is in the  $B_n$ theory. It is clear that the map is a bijection so that $X_S^{-1}$ is well defined.

\begin{figure}[!ht]
  \begin{center}
    \includegraphics[width=4in]{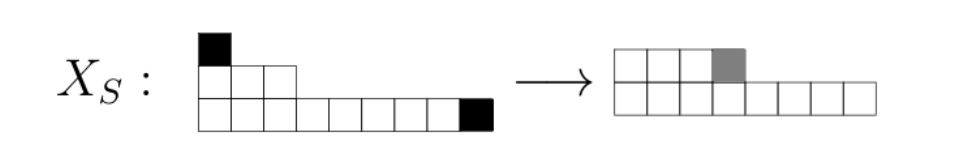}
  \end{center}
  \caption{The black boxes  of  the left Young tableau are  removed. And then they  are  putted at the end  of the  rows below,  denoted as  gray boxes. Under the  map  $X_s$, old rows  become   even rows. }
  \label{xs}
\end{figure}

The map $X_S$ can be described as removing one box from the end of the  $(2k+1)th$ row in the Young tableau and taking this box to the end of the $(2k)th$ row  as shown in Fig.\ref{xs}.

 Obviously,  this map preserves the rigidity conditions.   Note that the black boxes at the end of the  first row  disappear. So  the number of  boxes in the left Young tableau  is one more than that of  boxes in  the right Young tableau, which is consistent with the fact that the  right partition is in the $C_n$ theory.

The map $X_S$ preserves   symbol.  The   symbol invariant can be calculated by the definition
\begin{equation} \label{spesym}
 \Bigg(\!\!\!\begin{array}{c}  0\cdots 0\;\; \overbrace{1\cdots 1}^{n_2} \;\; 1 \dots 1 \;\; \cdots  \\
\;\;\;\; \underbrace{1 \cdots 1}_{n_1} \;\; 1\cdots 1 \;\;  \underbrace{2 \cdots 2}_{n_3}\;\;\cdots  \end{array} \!\!\!\Bigg)
\end{equation}
which is consistent with the result computed by  \textbf{Rule} \ref{rule} in the previous section. When $m=1$, this map is defined as
$$X_S: 1^{2n+1}\mapsto 1^{2n}.$$
 Acting on an empty set, the inverse map $X_S^{-1}$  give the partition '1'
$$X_S^{-1}: \emptyset \mapsto 1.$$

In \cite{Wyllard:2009}, Wyllard  proposed  an algorithm to construct dual of unipotent operators in the $B_n$ theory: Firstly, split the Young tableau $\rho$ into  tableau $\rho_{even}$ constructed from   even rows only  and one $\rho_{odd}$  constructed from the  old rows only.  Then take  the map $X_S$ turns $\rho_{odd}$ to a partition with only even rows.  While   $\rho_{even}$ is left  unchanged. Finally, the duality partition corresponding to $\rho$ in $C_n$ theory is $(X_S\rho_{odd}, \rho_{even})$.   This duality map is denoted by $WB$  preserving the symbol
\begin{equation}\label{SSB}
  WB:\,\,\,(\lambda, \emptyset)_B\rightarrow (\lambda_{old}+\lambda_{even},\emptyset)\rightarrow (X_S\lambda_{old},\lambda_{even})_C.
\end{equation}
An example illustrates this process.
\begin{flushleft}
\textbf{Example}: For the $B_{16}$ partition $5\,4^2\,3\,2^4\, 1$,  the Young tableau is
\end{flushleft}
\begin{equation}
\tableau{1 3 4 8 9}
\end{equation}
Let us split it into partitions  $\rho_{odd}$ and $\rho_{even}$
\begin{equation}\label{WB11}
\tableau{1 3 4 8 9} \rightarrow  \left(\tableau{1 3 9}\quad  ;\quad \tableau{4 8}\right).
\end{equation}
The map $X_S$  take  the  special $B_n$ partition $\rho_{odd}$ to a special $C_n$ partition
\begin{equation}
X_S:\,\,\,\tableau{1 3 9} \rightarrow  \tableau{4 8}.
\end{equation}
 After leaving the  partition  $\rho_{even}$  on the right side of  (\ref{WB11}) untouched,  we arrive at  a rigid semisimple  surface operator $(2^4\,1^4;2^4\,1^4)$ in the $C_n$ theory
 \begin{equation}
WB:\,\,\,\tableau{1 3 4 8 9}\rightarrow \left(\tableau{4 8} ; \tableau{4 8}\right).\non
\end{equation}
Using  \textbf{Rule} \ref{rule}, we have
 \begin{eqnarray*}
&&\sigma_{(2^4\,1^4;2^4\,1^4)}^{C}=\sigma_{(2^4\,1^4)}^{C}+\sigma_{(2^4\,1^4)}^{C}=\\
&&\left(\begin{array}{@{}c@{}c@{}c@{}c@{}c@{}c@{}c@{}c@{}c@{}} 0&&0&&0&&0&&0 \\ & 1 && 1 && 1 &&1& \end{array} \right) +
 \left(\begin{array}{@{}c@{}c@{}c@{}c@{}c@{}c@{}c@{}c@{}c@{}} 0&&0&&0&&1&&1 \\ & 0 && 0 &&0&&0 & \end{array} \right)+
 \left(\begin{array}{@{}c@{}c@{}c@{}c@{}c@{}c@{}c@{}c@{}c@{}} 0&&0&&0&&0&&0 \\ & 1 && 1 && 1 &&1& \end{array} \right) +
 \left(\begin{array}{@{}c@{}c@{}c@{}c@{}c@{}c@{}c@{}c@{}c@{}} 0&&0&&0&&1&&1 \\ & 0 && 0 &&0&&0 & \end{array} \right)
 \end{eqnarray*}
which is consistent with the symbol of the $B_{16}$ partition $5\,4^2\,3\,2^4\, 1$.

Summary, under the map $X_S$,  we have
\begin{equation}\label{B}
\sigma^B_{(\lambda)}=\sigma^C_{(X_S\lambda_{old}, \lambda_{even})}=\sigma^C_{(X_S\lambda_{old})}+\sigma^C_{(\lambda_{even})}.
\end{equation}
 The symbol  of the $B_n$ partition $\rho_{odd}$ is equal to that  of the image of the map $X_S$.
The contribution to symbol of $\lambda_{even}$  as even rows in the $B_n$ theory  is equal to that of $\lambda_{even}$ in the $C_n$ theory.

\subsection{Map $Y_S$}

Now we introduce another map $Y_S$ which map  the odd-row tableaux $\rho_{old}$ in the $C_n$ side to  a rigid $D_{n-k}$ partition $\rho_{even}$
\begin{eqnarray} \label{YS}
Y_S:&& m^{2n_m+1}\, (m-1)^{2n_{m-1}}\, (m-2)^{2n_{m-2}  } \cdots 2^{n_2} \, 1^{2n_1}  \non \\   & \mapsto&
m^{2n_m}\, (m-1)^{2n_{m-1} +2 }\, (m-2)^{2n_{m-2} - 2 } \cdots 2^{n_2-2} \, 1^{2n_1+2}\,
\end{eqnarray}
 where $m$ has to be even in order for the first element to be a $C_k$ partition. This map is a bijection  which takes a special $C_k$ partition to a special $D_k$ partition. It is clear that $Y_s$ preserves the number of boxes and  symbol.  A simple example illustrates this rule.
\begin{figure}[!ht]
  \begin{center}
    \includegraphics[width=4in]{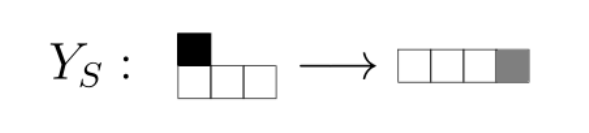}
  \end{center}
  \caption{The map $Y_S$ can be described as taking one box in the end of the  $(2m)$-th row  to the end of the   $(2m+1)$-th row. Under the map, old rows  become  even rows.
 }
  \label{ys}
\end{figure}

In \cite{Wyllard:2009}, Wyllard  proposed  an algorithm to construct dual of the unipotent operators on the $C_n$ side: Firstly, split the Young tableau $\rho$  into even-row tableau $\rho_{even}$   and odd-row $\rho_{odd}$  tableaux.  Then take  the map $X_{S}^{-1}$ turns the even-row tableau $\rho_{even}$ to a $B_k$ tableau and the map $Y_{S}$ turn the old-row tableau  $\rho$ to  a $D_k$ tableau.     This duality map is denoted by $WC$,
\begin{equation}\label{SSC}
 WC:\,\,\, (\lambda, \emptyset)_C \rightarrow (\lambda_{old}+\lambda_{even},\emptyset)\rightarrow (X_S^{-1}\lambda_{even},Y_S\lambda_{old})_B,
\end{equation}
which has been proved preserve the symbol invariant in \cite{Wyllard:2009}.

Summary, we have
\begin{equation*}
\sigma^C_{(\lambda)}=\sigma^B_{(X_S^{-1}\lambda_{even},Y_S\lambda_{old})}=\sigma^B_{(X_S^{-1}\lambda_{even})}+\sigma^D_{(Y_S\lambda_{old})}.
\end{equation*}
Since the maps $X_S$ preserve symbol,  the  above equation  reduces to
\begin{equation}\label{C}
\sigma^C_{(\lambda)}=\sigma^C_{(\lambda_{even})}+\sigma^D_{(Y_S\lambda_{old})}.
\end{equation}

For the unipotent surface operators in the  $D_n$ theory, Wyllard made the following proposal
\begin{equation}\label{WD0}
  WD:\,\,\, (\lambda, \emptyset)_D \rightarrow (\lambda_{\mathrm{old}}+\lambda_{\mathrm{even}},\emptyset)\rightarrow (\lambda_{\mathrm{even}},Y_S\lambda_{\mathrm{old}})_D.
\end{equation}
Start by splitting the corresponding tableau into even-and old-row tableaux,  leaving the even-row tableau unchanged and applying the map $Y_S$ to the odd-row tableau. The map $Y_{S}$ turn an old-row tableau into  an even-row tableau, which has been proved preserve the symbol invariant in \cite{Wyllard:2009}. Summary, we have
\begin{equation}\label{D}
\sigma^D_{(\lambda,\emptyset)}=\sigma^D_{(\lambda_{even})}+\sigma^D_{(Y_s\lambda_{old})}.
\end{equation}

\subsection{Calculating   symbol in a simple way }
From the above discussions, using the  formulas (\ref{B}), (\ref{C}), and (\ref{D}), a natural  construction of symbol of a rigid semisimple surface operator $(\lambda, \rho)$  in the   $B_n$,  $C_n$, and $D_n$   theories  now presents itself:
\begin{itemize}
  \item For  a rigid semisimple surface operator  $(\lambda, \rho)$  in the  $B_n$ theory, by using the formulas (\ref{B}) and (\ref{D}),  the symbol is
\begin{equation}\label{SB}
\sigma^B_{(\lambda,\rho)}=\sigma^B_{(\lambda)}+\sigma^D_{(\rho)}=\sigma^C_{(\lambda_{even})}+\sigma^C_{(X_S\lambda_{old})}+\sigma^D_{(\rho_{even})}+\sigma^D_{(Y_S\rho_{old})}.
\end{equation}
 The four terms on the right  can be calculated by  \textbf{Rule} \ref{rule},  since  the  partitions  $\lambda_{even}$, $X_S\lambda_{old}$, $\rho_{even}$, and $Y_S\rho_{old}$ only have  even rows.

  \item   For  a rigid semisimple surface operator  $(\lambda, \rho)$  in the  $C_n$ theory, by using the formulas (\ref{C}),   the symbol is
\begin{equation}\label{SC}
\sigma^C_{(\lambda,\rho)}=\sigma^C_{(\lambda)}+\sigma^C_{(\rho)}=\sigma^C_{(\lambda_{even})}+\sigma^D_{(Y_s\lambda_{old})}+\sigma^C_{(\rho_{even})}+\sigma^D_{(Y_S\rho_{old})}.
\end{equation}

  \item
 For  a rigid semisimple surface operator  $(\lambda, \rho)$  in the  $D_n$ theory, by using the formula(\ref{D}),  the symbol is
\begin{equation}\label{SD}
\sigma^D_{(\lambda,\rho)}=\sigma^D_{(\lambda)}+\sigma^D_{(\rho)}=\sigma^D_{(\lambda_{even})}+\sigma^D_{(Y_s\lambda_{old})}+\sigma^D_{(\rho_{even})}+\sigma^D_{(Y_S\rho_{old})}.
\end{equation}

\end{itemize}
Hence, for  a  rigid surface operator in the $B_n$, $C_n$, and $D_n$ theories,  the symbol  can be constructed by using the formulas (\ref{SB}), (\ref{SC}), and (\ref{SD}).

\section{Rigid surface operators in the $B_n$/$C_n$ theories}\label{BC}
In \cite{Wyllard:2009}, Wyllard made  explicit proposals for how the $S$-duality map should act on unipotent surface operators and certain subclasses of semisimple surface operators.  These proposals, with common  characteristics,  are given in the  first  subsection.  In the second and third subsections, we will make   new proposals for  certain subclasses of semisimple surface operators, with  evidences  provided.

\subsection{Proposals for $S$-duality maps for  surface operators}
The $S$-duality maps proposed   in \cite{Wyllard:2009} passed all consistency checks.
\label{unip}
\begin{flushleft}
\textbf{For rigid unipotent operators $(\lambda, \emptyset)$ of the $B_n$ theory}

The $S$-duality map is
\end{flushleft}
\begin{equation}\label{WB}
 WB:\,\,\,(\lambda, \emptyset)_B\rightarrow (\lambda_{old}+\lambda_{even},\emptyset)\rightarrow (X_S\lambda_{old},\lambda_{even})_C
\end{equation}
which is the map (\ref{SSB}).
\begin{flushleft}
  \textbf{Example}: For the $B_{16}$ partition,  $\la = 5\, 4^2 \, 3^3\, 2^4\, 1^3$, applying the map $WB$,  we find
\end{flushleft}
\begin{equation} \label{rank16}
\tableau{1 3 6 10 13} \quad\mapsto \left(\quad\tableau{4 12}\quad  ;\quad \tableau{6 10}\right)
\end{equation}
which leads to the semisimple $C_{16}$ surface operator $(2^4\,1^8\,,\,2^6\,1^4)$.

\begin{flushleft}
\textbf{For   surface operators $(1;\de)$ of the $B_n$ theory }

The $S$-duality map is
\end{flushleft}
\begin{equation}\label{WB1}
WB1:\,\,\,(1;\de)_B\rightarrow (1\,;\delta_{\mathrm{even}}+\delta_{\mathrm{odd}} )\rightarrow (X_S 1+Y_S^{-1}\delta_{\mathrm{even}}\,;\delta_{\mathrm{odd}} )_C.
\end{equation}
Split the partition $\de$ into even and odd rows.   Apply $Y_S^{-1}$ to the even-row tableau and leave the odd-row tableau unchanged.

We can prove that the map $WB1$ preserve the symbol. For the surface operator $(1;\de)_B$,  using the formula (\ref{SB}),  the symbol is
\begin{equation}\label{SSB1}
\sigma^B_{(1;\de)}=\sigma^B_{1}+\sigma^D_{(\de)}=\sigma^C_{(X_S 1)}+\sigma^D_{(\de_{even})}+\sigma^D_{(Y_S\de_{old})}=\sigma^D_{(\de_{even})}+\sigma^D_{(Y_S\de_{old})}.
\end{equation}
For the surface operator $(X_S 1+Y_S^{-1}\delta_{\mathrm{even}}\,;\delta_{\mathrm{odd}} )_C$, using the formula (\ref{SC}), the symbol is
\begin{equation*}
\sigma^C_{(X_S 1+Y_S^{-1}\delta_{\mathrm{even}}\,;\delta_{\mathrm{odd}} )}=\sigma^C_{(X_S 1+Y_S^{-1}\delta_{\mathrm{even}})}+\sigma^C_{(\delta_{\mathrm{odd}})}=\sigma^D_{(Y_s(Y_S^{-1}\delta_{\mathrm{even}}))}+\sigma^D_{(Y_S\delta_{old})}
\end{equation*}
which is equal to the formula (\ref{SSB1}).

\begin{flushleft}
\textbf{For rigid unipotent operators $(\lambda, \emptyset)$ of the $C_n$ theory}

The $S$-duality map is
\end{flushleft}
\begin{equation}\label{WC}
 WC:\,\,\, (\lambda, \emptyset)_C\rightarrow (\lambda_{old}+\lambda_{even},\emptyset)\rightarrow (X_S^{-1}\lambda_{even},Y_S\lambda_{old})_B,
\end{equation}
which is  the map (\ref{SSC}).

\begin{flushleft}
\textbf{For semisimple  surface operators $(\rho\,;\rho)$ of the  $C_n$  theory }

The $S$-duality map is
\end{flushleft}
\begin{equation}\label{WCC}
WCC:\,\,\,(\rho\,;\rho)_C\rightarrow(\rho_{\mathrm{even}} +\rho_{\mathrm{odd}} \,;\rho_{\mathrm{odd}} + \rho_{\mathrm{even}}) \rightarrow (\rho_{\mathrm{even}} + X_S^{-1}\rho_{\mathrm{even}}\,;\rho_{\mathrm{odd}} + Y_S\rho_{\mathrm{odd}})_B.
\end{equation}
The first step is to split two equal tableaux into even-row tableaux $\rho_{\mathrm{even}}$ and odd-row tableaux $\rho_{\mathrm{odd}}$. The second step is to  apply the map $X_S$  to one of the odd-row tableaux and apply the map $Y_S^{-1}$  to the even-row tableau in the other semisimple factor. Next add the altered and unaltered even-row tableaux to form one of the two partitions in a semisimple $B_n$ operator. Finally,  do the same  to the odd-row tableaux. For the resulting partition,  the first partition is a $B_k$ partition and the second factor is a $D_{n-k}$ partition.

We can prove that the map $WCC$ preserve the symbol.
For the surface operator $(\rho\,;\rho)_C$,  using the formula (\ref{SC}),  the symbol is
\begin{equation}\label{SSC1}
\sigma^C_{(\rho\,;\rho)}=\sigma^C_{(\rho)}+\sigma^C_{(\rho)}=\sigma^C_{(\rho_{even})}+\sigma^D_{(Y_s\rho_{old})}+\sigma^C_{(\rho_{even})}+\sigma^D_{(Y_S\rho_{old})}.
\end{equation}
For the surface operator $(\rho_{\mathrm{even}} + X_S^{-1}\rho_{\mathrm{even}}\,;\rho_{\mathrm{odd}} + Y_S\rho_{\mathrm{odd}})_B$, using the formula (\ref{SB}), the symbol is
\begin{eqnarray*}
\sigma^B_{(\rho_{\mathrm{even}} + X_S^{-1}\rho_{\mathrm{even}}\,;\rho_{\mathrm{odd}} + Y_S\rho_{\mathrm{odd}})_B}&=& \sigma^B_{(\rho_{\mathrm{even}} + X_S^{-1}\rho_{\mathrm{even}})}+\sigma^D_{(\rho_{\mathrm{odd}} + Y_S\rho_{\mathrm{odd}})}\\
&=& \sigma^C_{(\rho_{even})}+\sigma^C_{(X_S(X_S^{-1}\rho_{even}))}+\sigma^D_{(\rho_{old})}+\sigma^D_{(Y_S\rho_{old})}\\
&=& \sigma^C_{(\rho_{even})}+\sigma^C_{(\rho_{even})}+\sigma^D_{(Y_s\rho_{old})}+\sigma^D_{(Y_S\rho_{old})}
\end{eqnarray*}
which is equal to the formula (\ref{SSC1}).

\begin{flushleft}
\textbf{Example}: For the $C_{14}$ partition,  $(4\, 3^2\, 2\, 1^2\,;\,4\, 3^2\, 2\, 1^2)$, we have
\end{flushleft}
\begin{eqnarray}
\left( \tableau{1 3 4 6} \; ; \;\tableau{1 3 4 6} \right) &\mapsto& \left( \tableau{4 6} + \tableau{1 3 7}  \; ;  \; \tableau{1 3} + \tableau{4}\right) \non \\
&=& \left( \tableau{1 3 4 6 7} ; \tableau{1 3 4} \right)
\end{eqnarray}
which leads to a semisimple operator in the $B_n$ theory.

\subsection{$S$-duality maps for rigid semisimple operators in the $C_n$ theory }
 In this subsection,   we  discuss the possible forms of   $S$-duality maps of  rigid surface operators $(\lambda\,;\rho)$ in the $C_n$ theory and extend   maps  proposed in previous section,  with the same manipulation rule at the level of Young tableaux. Start by splitting the partition $\lambda$ into even-row and odd-row tableaux  and do the same for the partition $\rho$. The map $WC$ (\ref{WC}) fix the possible form of the  $S$-duality map of  rigid surface operator as follows
$$CB:\,\,\,(\lambda\,;\rho)_C\rightarrow(\lambda_{\mathrm{even}} +\lambda_{\mathrm{odd}} \,;\rho_{\mathrm{odd}} + \rho_{\mathrm{even}}) \rightarrow ( X_S^{-1}\lambda_{\mathrm{even}}+A \,;Y_S\lambda_{\mathrm{odd}} + B)_B,$$
where $A$ and $B$ denote the uncertain parts related to the partition $\rho$. Setting $\lambda=\rho$, comparing with the map $WCC$ (\ref{WCC}), we find that $A=\rho_{\mathrm{even}}$ and $B=\rho_{\mathrm{odd}}$. So the  $S$-duality map of  a general rigid semisimple surface operator $(\lambda\,;\rho)$ can be fixed as follows
\begin{equation}\label{CB}
CB:\,\,\,(\lambda\,;\rho)_C\rightarrow(\lambda_{\mathrm{even}} +\lambda_{\mathrm{odd}} \,;\rho_{\mathrm{odd}} + \rho_{\mathrm{even}}) \rightarrow (  X_S^{-1}\lambda_{\mathrm{even}}+\rho_{\mathrm{even}}\,; Y_S\lambda_{\mathrm{odd}}+\rho_{\mathrm{odd}} )_B.
\end{equation}
The inverse map $CB^{-1}$ for a rigid semisimple surface operator in the $B_n$ theory is
\begin{equation}\label{CBI}
CB^{-1}:\,\,\,(\lambda\,;\rho)_B\rightarrow( X_S\lambda_{\mathrm{old}}+ Y_S^{-1}\rho_{\mathrm{even}}\,;\lambda_{\mathrm{even}} +\rho_{\mathrm{old}} )_C.
\end{equation}
To be well defined,  the images of the maps $CB$ and $CB^{-1}$ should be  rigid semisimple surface operators.


\subsection{$S$-duality maps for rigid semisimple operators in the $B_n$ theory }
The map $WB$ (\ref{WB}) fix the possible forms of the  $S$-duality map of  rigid surface operators $(\lambda\,;\rho)$ in the $B_n$ theory as follows
$$BC:\,\,\,(\lambda\,;\rho)_B\rightarrow(\lambda_{\mathrm{even}} +\lambda_{\mathrm{odd}} \,;\rho_{\mathrm{odd}} + \rho_{\mathrm{even}}) \rightarrow ( X_S\lambda_{\mathrm{old}}+A \,;\lambda_{\mathrm{even}} + B)_C$$
where $A$ and $B$ denote the uncertain parts related to the partition $\rho$. Setting $\lambda=1$ and comparing with the map $WB1$ (\ref{WB1}), we find that there are two choices for the parts $A$ and $B$. The first one is  $A=Y_S^{-1}\rho_{\mathrm{even}}$ and $B=\rho_{\mathrm{old}}$ and the second one is $A=\rho_{\mathrm{old}}$ and $B=Y_S^{-1}\rho_{\mathrm{even}}$ .

For the first choice,  the  $S$-duality map  can be fixed as follows
\begin{equation}\label{BC1}
BC1:(\lambda;\rho)_B\rightarrow(\lambda_{\mathrm{even}} +\lambda_{\mathrm{odd}} ;\rho_{\mathrm{odd}} + \rho_{\mathrm{even}}) \rightarrow ( X_S\lambda_{\mathrm{old}}+ Y_S^{-1}\rho_{\mathrm{even}};\lambda_{\mathrm{even}} +\rho_{\mathrm{old}} )_C
\end{equation}
which is equal to  the map $CB^{-1}$ (\ref{CBI}).  To be well defined,  the images of the maps $CB$ and $CB^{-1}$ should be  rigid semisimple surface operators.


For the second choice,  the  $S$-duality map can be fixed as follows
\begin{equation}\label{BC2}
BC2:(\lambda;\rho)_B\rightarrow(\lambda_{\mathrm{even}} +\lambda_{\mathrm{odd}} ;\rho_{\mathrm{odd}} + \rho_{\mathrm{even}}) \rightarrow ( X_S\lambda_{\mathrm{old}}+\rho_{\mathrm{old}};\lambda_{\mathrm{even}} + Y_S^{-1}\rho_{\mathrm{even}})_C.
\end{equation}
However,  it  is not consistent with the map $CB$ (\ref{CB}) for all rigid partition pair $(\lambda\,;\rho)_B$. Applying the map  $CB$  to the rigid surface operator $(\lambda\,;\rho)=(\lambda, \emptyset)$ in the $C_n$ theory,  we have
$$CB:\, (\lambda, \emptyset)_C\rightarrow (\lambda_{\mathrm{old}}+\lambda_{\mathrm{even}},\emptyset)\rightarrow (X_S^{-1}\lambda_{\mathrm{even}},Y_S\lambda_{\mathrm{old}})_B.$$
Then applying the inverse map of  $CB$ to $(\lambda^{'}\,;\rho^{'})_B=(X_S^{-1}\lambda_{\mathrm{even}},Y_S\lambda_{\mathrm{old}})_B$, we have
$$BC2:\,\,\,(\lambda^{'}\,;\rho^{'})_B\rightarrow( X_S(X_S^{-1}\lambda_{\mathrm{even}})_{\mathrm{old}}\,;  Y_S^{-1}(Y_S\lambda_{\mathrm{old}})_{\mathrm{even}})=(\lambda_{\mathrm{even}}\,;  \lambda_{\mathrm{old}})_C$$
which is not equal to $ (\lambda, \emptyset)_C$. For certain subclasses rigid surface operators, we find  a  map $BC2$ which is   consistent with the map $CB$ in section \ref{other}.

To be well defined,  the images of all the $S$-duality maps  should be  rigid semisimple surface operators.

\subsection{$S$-duality maps for rigid surface operators in the $\SO(13)$ and $\Sp(12)$ theories}\label{tb}
In the following   table, we list  all rigid surface operators in the  $\Sp(12)$ and $\SO(13)$ theories to illustrate the proposed $S$-duality maps. The first column is the type of  the duality maps.  The second and third columns   list   pairs of partitions corresponding to the surface operators in the $B_n$ and $C_n$ theories. The other  columns are  the  dimension,   symbol invariant,   and  fingerprint invariant of the surface operator, respectively. Combining the  discussions in the following  subsections, all the  candidates  $S$-duality  pairs of   rigid surface operators   can  fit into the  proposal   duality relationships.  Even the   mismatch in the total number of rigid surface operators in the $B_n$ and $C_n$ theories can be explained.

\begin{equation} {
\begin{array}{l@{\hspace{10pt}}l@{\hspace{10pt}}l@{\hspace{10pt}}l@{\hspace{10pt}}l@{\hspace{10pt}}c@{\hspace{10pt}}l}
\underline{Num}
&
\underline{Type}
&
\underline{Sp(12)}
&
\underline{SO(13)}
&
\underline{Dim}
&
\underline{Symbol}
&
\underline{Fingerprint}
\\[-0.5pt]
1
&
CB
&
(1^{12}\,;\emp)
&
(1^{13};\emp)
 &
 0
 &
\left(\begin{array}{@{}c@{}c@{}c@{}c@{}c@{}c@{}c@{}c@{}c@{}c@{}c@{}c@{}c@{}c@{}}
 0&&0&&0&&0&&0&&0&&0 \\
 &1&&1&&1&&1&&1&&1&
\end{array}\right)
 &
[1^6;\emp]
 \\[-0.5pt]
 2
&
CB
&
 (2\,1^{10}\,;\emp)
 &
(1;1^{12})
 &
 12
 &
\left(\begin{array}{@{}c@{}c@{}c@{}@{}c@{}c@{}c@{}c@{}c@{}c@{}c@{}c@{}}
 1&&1&&1&&1&&1&&1 \\
 &0&&0&&0&&0&&0&
\end{array}\right)
 &
[1^5;1]
 \\[-0.5pt]
3
&
CB
&
  (1^{10}\,;1^2)
 &
(2^2\,1^9;\emp)
 &
 20
 &
\left(\begin{array}{@{}c@{}c@{}c@{}c@{}c@{}c@{}c@{}c@{}c@{}c@{}c@{}}
 0&&0&&0&&0&&0&&0 \\
 &1&&1&&1&&1&&2&
\end{array}\right)
 &
[2\,1^4;\emp]
 \\[-0.5pt]
 4
&
CB
&
 (2^3\,1^6\,;\emp)
 &
(1;2^2\,1^8)
 &
 30
 &
\left(\begin{array}{@{}c@{}c@{}c@{}@{}c@{}c@{}c@{}c@{}c@{}c@{}}
 1&&1&&1&&1&&1 \\
 &0&&0&&0&&1&
\end{array}\right)
 &
[1^3;1^3]
\\[-0.5pt]
5
&
CB_{eo}
&
(2\,1^8\,;1^2)
 &
(1^3;1^{10})
 &
 30
 &
\left(\begin{array}{@{}c@{}c@{}c@{}@{}c@{}c@{}c@{}c@{}c@{}c@{}c@{}c@{}}
 1&&1&&1&&1&&1 \\
 &0&&0&&0&&1&
\end{array}\right)
 &
[1^3;1^3]
 \\[-0.5pt]
 6
&
CB
&
 (1^{8}\,;1^4)
 &
(2^4\,1^5;\emp)
 &
 32
 &
\left(\begin{array}{@{}c@{}c@{}c@{}c@{}c@{}c@{}c@{}c@{}c@{}}
 0&&0&&0&&0&&0 \\
 &1&&1&&2&&2&
\end{array} \right)
 &
[2^2\,1^2;\emp]
 \\[-0.5pt]
7
&
CB
&
 (2^4\,1^4\,;\emp)
 &
(3\,2^2\,1^6;\emp)
 &
 36
 &
\left(\begin{array}{@{}c@{}c@{}c@{}c@{}c@{}c@{}c@{}c@{}c@{}}
 0&&0&&0&&1&&1 \\
 &1&&1&&1&&1&
\end{array}\right)
 &
[1^2;1^4]
 \\[-0.5pt]
 8
&
CB_{eo}
&
 (1^8\,;2\,1^2)
 &
(1^9,1^4)
 &
 36
 &
\left(\begin{array}{@{}c@{}c@{}c@{}c@{}c@{}c@{}c@{}c@{}c@{}}
 0&&0&&0&&1&&1 \\
 &1&&1&&1&&1&
\end{array}\right)
 &
[1^2;1^4]
 \\[-0.5pt]
9
&
CB
&
 (1^6\,;1^6)
 &
(2^6\,1;\emp)
 &
 36
 &
\left( \begin{array}{@{}c@{}c@{}c@{}c@{}c@{}c@{}c@{}}
 0&&0&&0&&0 \\
 &2&&2&&2&
\end{array} \right)
 & [2^3;\emp]
 \\[-0.5pt]
 10
&
CB
&
 (2^5\,1^2\,;\emp)
 &
(1;2^4\,1^4)
 &
 40
 &
\left(\begin{array}{@{}c@{}c@{}c@{}@{}c@{}c@{}c@{}c@{}}
 1&&1&&1&&1 \\
 &0&&1&&1&
\end{array}\right)
 &
[1;1^5]
 \\[-0.5pt]
 11
&
CB_{eo}
&
  (2\,1^6\,;1^4)
 &
(1^5;1^8)
 &
 40
 &
\left(\begin{array}{@{}c@{}c@{}c@{}@{}c@{}c@{}c@{}c@{}c@{}c@{}}
 1&&1&&1&&1 \\
 &0&&1&&1&
\end{array}\right)
 &
[1;1^5]
 \\[-0.5pt]
 12
&
CB_{eo}
&
  (1^6\,;2\,1^4)
 &
(1^7;1^6)
 &
 42
 &
\left(\begin{array}{@{}c@{}c@{}c@{}c@{}c@{}c@{}c@{}}
 0&&1&&1&&1 \\
 &1&&1&&1&
\end{array}\right)
 &
[\emp;1^6]
 \\[-0.5pt]
 13
&
CB
&
 (3^2\,2\,1^4\,;\emp)
 &
(1^3;2^2\,1^6)
 &
 44
 &
\left(\begin{array}{@{}c@{}c@{}c@{}@{}c@{}c@{}c@{}c@{}c@{}c@{}}
 1&&1&&1&&1 \\
 &0&&0&&2&
\end{array}\right)
 &
[3\,1^2;1]
 \\[-0.5pt]
 14
&
N1
&
  (2^3\,1^4\,;1^2)
 &
(2^2\,1;1^8)
 &
 44
 &
\left(\begin{array}{@{}c@{}c@{}c@{}@{}c@{}c@{}c@{}c@{}c@{}c@{}}
 1&&1&&1&&1 \\
 &0&&0&&2&
\end{array}\right)
 &
[3\,1^2;1]
 \\[-0.5pt]
 15
&
CB
&
(2\,1^6\,;2\,1^2)
 &
(1;3\,2^2\,1^5)
 &
 44
 &
\left(\begin{array}{@{}c@{}c@{}c@{}@{}c@{}c@{}c@{}c@{}}
 1&&1&&2&&2 \\
 &0&&0&&0&
\end{array}\right)
 &
[2\,1^2;2]
 \\[-0.5pt]
 16
&
N2
&
  (2^4\,1^2\,;1^2)
 &
(2^21^5;1^4)
 &
 48
 &
\left(\begin{array}{@{}c@{}c@{}c@{}c@{}c@{}c@{}c@{}}
 0&&0&&1&&1 \\
 &1&&1&&2&
\end{array}\right)
 &
[3\,1;1^2]
 \\[-0.5pt]
 17
&
CB
 &
  (2\,1^4\,;2\,1^4)
 &
(1;3\,2^4\,1)
 &
 48
 &
\left(\begin{array}{@{}c@{}c@{}c@{}@{}c@{}c@{}}
 2&&2&&2 \\
 &0&&0&
\end{array}\right)
 &
[2^2;2]
 \\[-0.5pt]
 18
&
CB_{eo}
&
  (2^3\,1^2\,;1^4)
 &
(1^5;2^2\,1^4)
 &
 50
 &
\left(\begin{array}{@{}c@{}c@{}c@{}c@{}c@{}}
  1&&1&&1 \\
 &1&&2&
\end{array}\right)
 &
[3;1^3]
 \\[-0.5pt]
 19
&
-
&
-
 &
 (2^21^3;1^6)
&
 50
 &
\left(\begin{array}{@{}c@{}c@{}c@{}@{}c@{}c@{}c@{}c@{}}
 1&&1&&1 \\
 &1&&2&
\end{array}\right)
 &
[3;1^3]
 \\[-0.5pt]
 20
&
-
&
-
 &
(2^4\,1;1^4)
 &
 52
 &
\left(\begin{array}{@{}c@{}c@{}c@{}c@{}c@{}}
 0&&1&&1 \\
 &2&&2&
\end{array}\right)
 &
[3^2;\emp]
\end{array}
}\non
\end{equation}

\begin{equation} {
\begin{array}{l@{\hspace{10pt}}l@{\hspace{10pt}}l@{\hspace{10pt}}l@{\hspace{10pt}}l@{\hspace{10pt}}c@{\hspace{10pt}}l}
\underline{Num}
&
\underline{Type}
&
\underline{Sp(12)}
&
\underline{SO(13)}
&
\underline{Dim}
&
\underline{Symbol}
&
\underline{Fingerprint}
\\[-0.5pt]
 21
&
N3
&
 (2^3\,1^2\,;2\,1^2)
 &
(1^3;3\,2^2\,1^3)
 &
 54
 &
\left(\begin{array}{@{}c@{}c@{}c@{}@{}c@{}c@{}c@{}c@{}}
  1&&2&&2 \\
  &0&&1&
\end{array}\right)
 &
[3\,1;2]
\\[-0.5pt]
22^{*}
&
CB
&
 (3^2\,2\,1^2\,;1^2)
&
(2^2\,1;2^2 \,1^4)
 &
 54
 &
\left(\begin{array}{@{}c@{}c@{}c@{}@{}c@{}c@{}c@{}c@{}}
 1&&1&&1 \\
 &0&&3&
\end{array}\right)
 &
[4\,1;1]
 \\[-0.5pt]
23
&
-
&
-
 &
(1^5;3\,2^2\,1)
 &
 56
 &
\left(\begin{array}{@{}c@{}c@{}c@{}@{}c@{}c@{}c@{}c@{}}
 0&&2&&2 \\
 &1&&1&
\end{array}\right)
 &
[3;2\,1]
 \\[-0.5pt]
24
&
-
&
-
 &
(2^2\,1;3\,2^2\,1)
 &
 60
 &
\left(\begin{array}{@{}c@{}c@{}c@{}}
 2&&2 \\
 &2&
\end{array}\right)
 &
[\emp;2^3]
\end{array}
}\non
\end{equation}

\subsection{Other $S$-duality maps}\label{other}
To be a well defined  $S$-duality map, the image of $BC1$ should be   rigid semisimple surface operators, with rigid partitions $X_S\lambda_{\mathrm{old}}$  and $Y_S^{-1}\rho_{\mathrm{even}}$ in the  $C_n$ theory.
Applying the map $BC1$ to the rigid surface operator $(1^{2m+1}\,; 1^{2n})$, we have the surface operator $(X_S1^{2m+1}+Y_S^{-1}1^{2n}\,; \emptyset)$. For $n\geq m$, the resulting partition  $32^{2m-1}1^{2(n-m)}$ is not a rigid partition in the  $C_n$ theory. For  $n\leq m$, we draw the same conclusion.

However, there is a  $S$-duality pair in the table of previous subsection
\begin{equation}\label{S}
S:\,\,\, (1^{2m+1}\,;1^{2n})_B\rightarrow(1^{2m}\,;21^{2(n-1)})_C.
\end{equation}
We can prove that  the map $S$ preserve the symbol invariant  as well as dimension
$$d_{(1^{2m+1}\,;1^{2n})_B}=d_{(1^{2m}\,;21^{2(n-1)})_C}.$$
It suggest us to propose the  following  $S$-duality map
\begin{equation}\label{BCeo}
  BC_{oe}:\,\,\,(\lambda_{\mathrm{old}}\,;\rho_{\mathrm{even}})_B\rightarrow(X_S\lambda_{\mathrm{old}}\,;Y_S^{-1}\rho_{\mathrm{even}})_C
\end{equation}
By using the map $X_S$, the partition $1^{2m+1}$ in the $B_n$ theory become  a partition  $1^{2m}$ in the $C_n$ theory.
By using the map $Y_S^{-1}$, the partition $1^{2n}$ in the $D_n$ theory become  the partition  $21^{2(n-1)}$ in the $C_n$ theory.

The map $BC_{oe}$ is a special case of the map $BC2$  (\ref{BC2}), whose inverse map can be proposed as follows
$$CB_{eo}:\,\,\,(\lambda_{\mathrm{even}}\,;\rho_{\mathrm{old}})_C\rightarrow(X_S^{-1}\lambda_{\mathrm{even}}\,;Y_S\rho_{\mathrm{old}})_B.$$
The following   example of  the map $BC2$ in the preceding table  is  not included in the map $S$ (\ref{S}).
\begin{flushleft}
  \textbf{Example}: The 18th  duality pair
   $$CB_{eo}:\,\,\,  (2^3\,1^2\,;1^4)_C\rightarrow(1^5;2^2\,1^4)_B.$$
\end{flushleft}

\subsection{Exceptions}
There are several $S$-duality pairs in the table of section \ref{tb}   not belonging  to any proposed $S$-duality maps.  We discuss   $S$-duality maps of the 14th, 16th, 19th, and  21th  pairs independently.

\begin{flushleft}
\textbf{The 14th  duality pair}: we have the map $N1:\,\,(2^2\,1\,;1^8)_B\rightarrow(2^3\,1^4\,;1^2)_C$.
\end{flushleft}
\begin{eqnarray}
\left( \tableau{2 3} \; ; \;\tableau{8} \right) &\mapsto& \left( \tableau{2} + \tableau{3}  \; ;  \; \tableau{8}\right) \non \\
&\rightarrow & \left( Y_S^{-1}(X_S\tableau{3} + \tableau{8})  \; ;  \; \tableau{2}\right) \non \\
&=& \left( \tableau{3 7} \; ; \tableau{2} \right)
\end{eqnarray}

\begin{flushleft}
\textbf{The 16th  duality pair}: we have the map $N2:\,\,(2^2\,1^5\,;1^4)_B\rightarrow(2^4\,1^2\,;1^2)_C$.
\end{flushleft}
\begin{eqnarray}
\left( \tableau{2 7} \; ; \;\tableau{4} \right) &\mapsto& \left( \tableau{2} + \tableau{7}  \; ;  \; \tableau{4}\right) \non \\
&\rightarrow & \left( X_S\tableau{7} + \tableau{4}  \; ;  \; \tableau{2}\right) \non \\
&=& \left( \tableau{4 6} \; ; \tableau{2} \right)
\end{eqnarray}

\begin{flushleft}
\textbf{The 21th duality pair}: we have the map $N3: (1^3\,;3\,2^2\,1^3)_B\rightarrow(2^3\,1^2\,;2\,1^2)_C $.
\end{flushleft}
\begin{eqnarray}
\left( \tableau{3} \; ; \;\tableau{1 3 6} \right) &\mapsto& \left( \tableau{3}  \; ;  \;  \tableau{1 3}+\tableau{6}\right) \non \\
&\rightarrow & \left( Y_S^{-1}(X_S\tableau{3} + \tableau{6})  \; ;  \; \tableau{1 3}\right) \non \\
&=& \left( \tableau{3 5} \; ; \tableau{1 3} \right)
\end{eqnarray}

For the  19th  duality pair, the $B_n$ surface operator $(1^3\,;3\,2^2\,1^3)_B$ do not have candidate duals. Since there is only one $C_n$ rigid surface operator $(2^3\,1^2\,;2\,1^2)_C $ have the same invariants with it,  we  propose the following $S$-duality map artificially \footnote{It imply that there should be a mechanism for mapping several rigid surface operators with the same invariants in the $B_n$ theory to a rigid surface operator in the $C_n$ theory. }.
\begin{flushleft}
\textbf{The 19th  duality pair}: we have the map $N:\,\,(2^2\,1^3\,;1^6)_B\rightarrow(2^3\,1^2\,;2\,1^2)_C $.
\end{flushleft}
\begin{eqnarray}
\left( \tableau{2 5} \; ; \;\tableau{6} \right) &\mapsto& \left( \tableau{3 5} \; ;  \; \tableau{4}\right) \non \\
&\rightarrow & \left(Y_S^{-1}( \tableau{2} + \tableau{6} ) \; ;  \; X_S\tableau{5}\right) \non \\
&=& \left( \tableau{3 5} \; ; \tableau{4} \right)
\end{eqnarray}

For the  20th, 23th, and 24th $S$-duality pairs, the $B_n$ surface operators  do not have candidate duals. In fact, for these operators,   we could not find the  $S$-duality surface operators in the $C_n$ theory   by  splitting   the factors of the rigid semisimple surface operator in the $B_n$ theory to an even-row tableau and an old-row tableau, and then  applying   maps $X_S$ and $Y_S^{-1}$ to them.

The above  examples imply the  following conjecture.
\begin{flushleft}
\textbf{Conjecture:}
All the rigid surface operators in the $C_n$ theories can be obtained through the manipulations of splitting  the semisimple factors of a rigid surface operator in the $B_n$ theories into even tableaux and old tableaux,  the  actions of $X_S$  on part of old tableaux and $Y_S^{-1}$ on part of  the  even tableaux,  and combinations of  these building blocks  if necessary \footnote{A rigid surface operator in the $B_n$ theories do not have candidate duals if it can not leads to  a rigid $C_n$ surface operator through these manipulations rules. This conjecture implies the   mismatch on the total number of rigid surface operators in the $B_n$ and $C_n$ theories. }.
\end{flushleft}

\section{Rigid surface operators in the $D_n$ theory}\label{Ds}
Since the Langlands dual groups of  $D_n(SO(2n))$ are themselves, applying the $S$-duality map to a rigid semisimple surface operator, we would obtain a surface operator in the same theory.
For  unipotent surface operators, Wyllard made the following $S$-duality map  in \cite{Wyllard:2009}
\begin{equation}\label{WD}
  WD:\,\,\, (\lambda, \emptyset)_D \rightarrow (\lambda_{\mathrm{old}}+\lambda_{\mathrm{even}},\emptyset)\rightarrow (\lambda_{\mathrm{even}},Y_S\lambda_{\mathrm{old}})_D.
\end{equation}
which is the map (\ref{WD0}).

For the semisimple rigid $D_n$ surface operators of the form $(\rho\,;\rho)$, Wyllard made the following proposal in \cite{Wyllard:2009}
\begin{equation}\label{WDD}
 WDD:\,\,\,(\rho\,;\rho)_D\rightarrow(\rho_{\mathrm{even}} +\rho_{\mathrm{odd}} \,;\rho_{\mathrm{odd}} + \rho_{\mathrm{even}}) \rightarrow (\rho_{\mathrm{even}} + Y_S^{-1}\rho_{\mathrm{even}}\,;Y_S\rho_{\mathrm{odd}}+\rho_{\mathrm{odd}})_D.
\end{equation}
Split the two equal tableaux into even-row and odd-row tableaux. Next apply  $Y_S^{-1}$ to one of the even-row tableau and $Y_S$ to one of the odd-row tableau. Then add the unchanged even-row tableau and the transformed even-row tableau. Do the same for the odd-row tableau.

In this section,  we will extend above maps  to a general  rigid semisimple  surface operator $(\lambda\,;\rho)$,  with the same manipulation rule at the level of Young tableaux. Start by splitting the partition $\lambda$ into even-row and odd-row tableaux  and do the same for the partition $\rho$. The map $WD$ fix the possible form of the  $S$-duality map  as follows
$$DD:\,\,\,(\lambda\,;\rho)_D\rightarrow(\lambda_{\mathrm{even}} +\lambda_{\mathrm{odd}} \,;\rho_{\mathrm{odd}} + \rho_{\mathrm{even}}) \rightarrow (\lambda_{\mathrm{even}} +A\,;Y_S\lambda_{\mathrm{odd}} +B )_D$$
where $A$ and $B$ denote the uncertain parts related to the partition $\rho$. Setting $\lambda=\rho$  and  comparing with the map $WDD$, we find that $A=Y_S^{-1}\rho_{\mathrm{even}}$ and $B=\rho_{\mathrm{odd}}$. So the  $S$-duality map of  a general rigid semisimple surface operator $(\lambda\,;\rho)$ is fixed as follows
\begin{equation}\label{DD}
  DD:\,\,\,(\lambda\,;\rho)_D\rightarrow(\lambda_{\mathrm{even}} +\lambda_{\mathrm{odd}} \,;\rho_{\mathrm{odd}} + \rho_{\mathrm{even}}) \rightarrow (\lambda_{\mathrm{even}} + Y_S^{-1}\rho_{\mathrm{even}}\,;\rho_{\mathrm{odd}} + Y_S\lambda_{\mathrm{odd}})_D.
\end{equation}
The inverse map of $DD$ is
$$DD^{-1}:(\lambda\,;\rho)_D \rightarrow (\lambda_{\mathrm{even}} +\lambda_{\mathrm{odd}} \,;\rho_{\mathrm{odd}} + \rho_{\mathrm{even}})\rightarrow (\lambda_{\mathrm{even}} + Y_S^{-1}\rho_{\mathrm{even}}\,;\rho_{\mathrm{odd}} + Y_S\lambda_{\mathrm{odd}})_D.$$
which is equal to the map $DD$ since  the $D_n$ theories are self-dual.

To illustrate the results in this section, we list  all the  $S$-duality pairs of rigid surface operators in the  $\SO(12)$ theories.
\begin{equation} {
\begin{array}{l@{\hspace{10pt}}l@{\hspace{15pt}}l@{\hspace{15pt}}l@{\hspace{15pt}}l@{\hspace{15pt}}c@{\hspace{15pt}}l}
\underline{Num}
&
\underline{Type}
&
\underline{SO(12)}
&
\underline{SO(12)}
&
\underline{Dim}
&
\underline{Symbol}
&
\underline{Fingerprint}
\\[-0.5pt]
1
&
WD
&
(1^{12}\,;\emp)
&
(1^{12};\emp)
 &
 0
 &
\left(\begin{array}{@{}c@{}c@{}c@{}@{}c@{}c@{}c@{}c@{}c@{}c@{}c@{}c@{}}
 1&&1&&1&&1&&1&&1 \\
 &0&&0&&0&&0&&0&
\end{array}\right)
 &
[1^6;\emp]
 \\[-0.5pt]
 2
&
WD
&
  (2^2\,1^8;\emp)
 &
(2^2\,1^8;\emp)
 &
 16
 &
\left(\begin{array}{@{}c@{}c@{}c@{}c@{}c@{}c@{}c@{}c@{}c@{}}
 1&&1&&1&&1&&1 \\
 &0&&0&&0&&1&
\end{array}\right)
 &
[2\,1^4;\emp]
 \\[-0.5pt]
 3
&
WD
&
 (2^4\,1^4;\emp)
 &
(2^4\,1^4;\emp)
 &
24
 &
\left(\begin{array}{@{}c@{}c@{}c@{}c@{}c@{}c@{}c@{}}
 1&&1&&1&&1 \\
 &0&&1&&1&
\end{array}\right)
 &
[2^2\,1^2;\emp]
 \\[-0.5pt]
 4
&
WD
&
(1^8,1^4)
 &
(3\,2^2\,1^5;\emp)
 &
 32
 &
\left(\begin{array}{@{}c@{}c@{}c@{}c@{}c@{}c@{}c@{}}
 1&&1&&2&&2 \\
 &0&&0&&0&
\end{array}\right)
 &
[1^2;1^4]
 \\[-0.5pt]
5
&
WDD
&
  (3\,2^4\,1;\emptyset)
 &
(1^6;1^6)
 &
36
 &
\left(\begin{array}{@{}c@{}c@{}c@{}@{}c@{}c@{}}
 2&&2&&2 \\
 &0&&0&
\end{array}\right)
 &
[\emp;1^6]
 \\[-0.5pt]
 6
&
D_{even}
&
  (2^21^4;1^4)
 &
(2^21^4;1^4)
 &
 40
 &
\left(\begin{array}{@{}c@{}c@{}c@{}@{}c@{}c@{}}
 1&&2&&2 \\
 &0&&1&
\end{array}\right)
 &
[3\,1;1^2]
 \\[-0.5pt]
7
&
DD
&
(3\,2^2\,1;1^4)
 &
(3\,2^2\,1;1^4)
 &
48
 &
\left(\begin{array}{@{}c@{}c@{}c@{}}
 3&&3 \\
 &0&
\end{array}\right)
 &
[3;2\,1]
 \\
\end{array}
}\non \end{equation}

The first three  rigid surface operators pairs  are self-duality under the map $WD$ (\ref{WD}).  The 4th  $S$-duality pair of rigid surface operators also belong to the map $WD$. The 5th   pair of rigid surface operators  belong to the map $WDD$ (\ref{WDD}).  While the 7th pair belongs to the extended map $DD$ (\ref{DD})
\begin{eqnarray*}
DD: \,\,\,(32^21;1^4) \rightarrow (32^21;1^4)
\end{eqnarray*}
 which is not proposed  by Wyllard \cite{Wyllard:2009}.

For the 6th rigid surface operators pairs,  we can not obtain a rigid semisimple $D_n$ surface operator by  applying the map $DD$  or other manipulations to the rigid surface operator $(2^21^4;1^4)$. With only one rigid surface operator of dimension  $40$, it must be self-duality. So we propose the following  $S$-duality  map for this rigid surface operator
$$D_{\mathrm{even}}:\,\,\,(\lambda_{\mathrm{even}}\,;\rho_{\mathrm{even}})\rightarrow (\lambda_{\mathrm{even}}\,;\rho_{\mathrm{even}}).$$

\section{Summary and open problems}
We have found simple rules to construct the symbol of a partition with only even rows.  Using these  calculation rules  and the maps $X_S$,  $Y_S$, we have simplified  the computation  of {\it symbols } for  the partitions in the $B_n$, $C_n$, and $D_n$ theories. These calculation  rules  of symbol are very convenient to search $S$-duality maps. By consistency checks,  we have recovered and extended   the $S$-duality maps proposed by Wyllard \cite{Wyllard:2009}. We have also found a  new subclasses of  rigid  surface operators related by $S$-duality. We tried  to explain the  exceptions of  $S$-duality maps and pointed out   some  common characteristics  for the surface operators related by $S$-duality.  The above techniques used in the $B_n/C_n$ theories can be extended  to the $D_n$ theories.

In addition to the symbol invariant,  we should continue the analysis by computing other invariants of surface operators, such as the fingerprint invariant, center and topology of these proposals made in \cite{Wyllard:2009} for complement. The computation of the fingerprint invariant is complicated,  which we  hope to  find rules to  simplify. Simple and even explicit formulas of the fingerprint and  the symbol are needed.

As shown in the table of the surface operators  in the $\SO(13)$ and $\Sp(12)$ theories, there are still several candidates duality  pairs of   rigid surface operators can not  fit into the proposed  dual maps    in this paper.  Even more serious,   there is a  mismatch on the total number of rigid surface operators in the $B_n$ and $C_n$ theories.  The physical reason for the mismatch is still unknown.  Maybe we  should also take account of the  weakly rigid surface operators discussed in \cite{Gukov:2008} or the quantum  effect to resolve  the discrepancy issue. Clearly more work is required since above reasons.

\acknowledgments
We would like to thank Xian Gao  and Song He for  many helpful discussions.  This work was supported by a grant from  the Postdoctoral Foundation of Zhejiang Province.

\appendix
\section{Summary relevant facts}

\subsection{Rigid Partitions in the $B_n$, $C_n$, and $D_n$ theories}
Firstly, we  prove  that the longest row of a partition  always contains an odd number of boxes  in the $B_n$ theory. For a rigid partition,  even integers appear an even number of times, so the sum of all old integers is old which  means the  number of old integers is old. It is imply that the length of the partition, which  is the sum of the number of old integers and even integers,  is old. So the longest row    contains an odd number of boxes.

Next, We  prove  that the following two rows of the first row are either both of odd length or both of even length,  this pairwise pattern then continues. Assuming these facts  are true for the first $2i+1$ rows of a partition, we  prove that the  next two rows are either both  odd length or both  even length.  Assuming that the $(2i+2)$-th row is old with length $2m+1$, if the $(2i+3)$-th row is even with length $2n$ then there are  $(2m+1)-(2n) $  integers $2i+2$ in the partition. It  is a contradictory to the fact that even integers appear an even number of times in a rigid $B_n$ partition. So the $(2i+3)$-th row is old. Using the same method, we can prove that  the $(2i+3)$-th row is even if the ($2i+2$)-th row is even.

Similarly, we can prove that the longest two rows in a rigid $C_n$ partition both contain either  an even or an odd number  number of boxes. This pairwise pattern then continues.

We can also prove that the longest row in a rigid $D_n$ partition always contains an even number of boxes. And the following two rows are either both of even length or both of old length. This pairwise pattern then continue.

\subsection{Symbol of  the first row of a  partition   in the $C_n$  and  $D_n$ theories}
According to the  \textbf{Rule} \ref{rule} in section \ref{inv},  the contribution to symbol of the  first even  row of a partition in the $C_n$, $ D_n$ theories is needed  as an initial conditions.  For the first row with  $2l$ boxes of a partition in the $C_n$ theory, we have
$$
\begin{array}{rrrrr}
                 \lambda_{k}  & 1  &\cdots & 1& 0\\
  2l+1-k:          & 2l &\cdots &1 & 0\\
 2l+1-k+\lambda_{l-k+1}: & 2l+1 &\cdots & 2 & 0
\end{array}
$$
then
$$
\begin{array}{rrrr|rrrrrrr}
 2f_i+1: & 2l+1 &\cdots & 3  & & & 2g_i :& 2l &\cdots & 2 & 0 \\
 f_i: & l &\cdots & 1       & & &g_i :& l &\cdots & 1 & 0
\end{array}
$$
Finally,  we get $\alpha_i=g_i-i+1 : 0,\cdots , 0 , 0 $ and $ \beta_i=f_i-i+1 : 1,\cdots ,1 , 0$. So the contribution to symbol of the first row is
\begin{equation}
 \Bigg(\!\!\!\begin{array}{c} 0\;\;0\cdots 0  \\
 \;\;\;\underbrace{1 \cdots 1}_{l} \ \end{array} \!\!\!\Bigg).
\end{equation}

Next, we  check the \textbf{Rule} \ref{rule} for the $C_n$ theory. We add a  row with $2m$  boxes to  the   first row, then   the partition is $2^{2m}1^{2l-2m}$.  The contribution to symbol of the second row is
\begin{equation}
 \Bigg(\!\!\!\begin{array}{c}0\;\;0\cdots 0\;\; \overbrace{1\cdots 1}^{m}  \\
\;\;\;0\cdots 0 \;\;0 \cdots 0\ \end{array} \Bigg).
\end{equation}
If  two   rows  with lengths of  $2m$ and $2n$ ($m>n$) boxes are added to the first row,  the contribution to symbol  is
\begin{equation}
 \Bigg(\!\!\!\begin{array}{c}0\;\;0\cdots 0 \;\; \overbrace{1\cdots 1\;\; 1\cdots1}^{m} \\
\;\;\;0\cdots 0 \;\;0\cdots0\;\underbrace{1\cdots 1}_{n} \ \end{array} \Bigg)
\end{equation}
which is consistent with the \textbf{Rule} \ref{rule}.

Compared  with the $B_n$ case, the   terms $g_i$ and $f_i$ reverse roles  for the calculation of symbol in  the  $D_n$ theory. So the first row with  $2l$ boxes contribute $l$  '1' in another row of the symbol
\begin{equation} \label{d1}
 \Bigg(\!\!\!\begin{array}{c} \overbrace{1\;\;1\cdots 1}^{l}  \\
 \;\;\;\underbrace{0 \cdots 0}_{l-1} \ \end{array} \!\!\!\Bigg).
\end{equation}


\end{document}